\documentclass{article}
\usepackage[margin=1in]{geometry}

\usepackage{amsmath}
\usepackage{url}
\usepackage{microtype}

\makeatletter
\@ifundefined{lhs2tex.lhs2tex.sty.read}%
  {\@namedef{lhs2tex.lhs2tex.sty.read}{}%
   \newcommand\SkipToFmtEnd{}%
   \newcommand\EndFmtInput{}%
   \long\def\SkipToFmtEnd#1\EndFmtInput{}%
  }\SkipToFmtEnd

\newcommand\ReadOnlyOnce[1]{\@ifundefined{#1}{\@namedef{#1}{}}\SkipToFmtEnd}
\usepackage{amstext}
\usepackage{amssymb}
\usepackage{stmaryrd}
\DeclareFontFamily{OT1}{cmtex}{}
\DeclareFontShape{OT1}{cmtex}{m}{n}
  {<5><6><7><8>cmtex8
   <9>cmtex9
   <10><10.95><12><14.4><17.28><20.74><24.88>cmtex10}{}
\DeclareFontShape{OT1}{cmtex}{m}{it}
  {<-> ssub * cmtt/m/it}{}

\DeclareFontShape{OT1}{cmtt}{bx}{n}
  {<5><6><7><8>cmtt8
   <9>cmbtt9
   <10><10.95><12><14.4><17.28><20.74><24.88>cmbtt10}{}
\DeclareFontShape{OT1}{cmtex}{bx}{n}
  {<-> ssub * cmtt/bx/n}{}

\newcommand{\Conid}[1]{\mathit{#1}}
\newcommand{\Varid}[1]{\mathit{#1}}
\newcommand{\anonymous}{\kern0.06em \vbox{\hrule\@width.5em}}

\renewcommand{\geq}{\geqslant}
\usepackage{polytable}

\@ifundefined{mathindent}%
  {\newdimen\mathindent\mathindent\leftmargini}%
  {}%

\def\resethooks{%
  \global\let\SaveRestoreHook\empty
  \global\let\ColumnHook\empty}
\newcommand*{\savecolumns}[1][default]%
  {\g@addto@macro\SaveRestoreHook{\savecolumns[#1]}}
\newcommand*{\restorecolumns}[1][default]%
  {\g@addto@macro\SaveRestoreHook{\restorecolumns[#1]}}
\newcommand*{\aligncolumn}[2]%
  {\g@addto@macro\ColumnHook{\column{#1}{#2}}}

\resethooks

\newcommand{\onelinecommentchars}{\quad-{}- }
\newcommand{\commentbeginchars}{\enskip\{-}
\newcommand{\commentendchars}{-\}\enskip}

\newcommand{\visiblecomments}{%
  \let\onelinecomment=\onelinecommentchars
  \let\commentbegin=\commentbeginchars
  \let\commentend=\commentendchars}

\newcommand{\invisiblecomments}{%
  \let\onelinecomment=\empty
  \let\commentbegin=\empty
  \let\commentend=\empty}

\visiblecomments

\newlength{\blanklineskip}
\newcommand{\hsindent}[1]{\quad}
\let\hspre\empty
\let\hspost\empty

\EndFmtInput
\makeatother
\ReadOnlyOnce{polycode.fmt}%
\makeatletter

\newcommand{\hsnewpar}[1]%
  {{\parskip=0pt\parindent=0pt\par\vskip #1\noindent}}

\newcommand{\hscodestyle}{}

\newcommand{\sethscode}[1]%
  {\expandafter\let\expandafter\hscode\csname #1\endcsname
   \expandafter\let\expandafter\endhscode\csname end#1\endcsname}

  {\par\noindent
   \advance\leftskip\mathindent
   \hscodestyle
   \let\\=\@normalcr
   \let\hspre\(\let\hspost\)%
   \pboxed}%
  {\endpboxed\)%
   \par\noindent
   \ignorespacesafterend}

  {\hsnewpar\abovedisplayskip
   \advance\leftskip\mathindent
   \hscodestyle
   \let\hspre\(\let\hspost\)%
   \pboxed}%
  {\endpboxed%
   \hsnewpar\belowdisplayskip
   \ignorespacesafterend}

  {\hsnewpar\abovedisplayskip
   \advance\leftskip\mathindent
   \hscodestyle
   \let\\=\@normalcr
   \(\pboxed}%
  {\endpboxed\)%
   \hsnewpar\belowdisplayskip
   \ignorespacesafterend}

\newcommand{\plainhs}{\sethscode{plainhscode}}

\plainhs

  {\hsnewpar\abovedisplayskip
   \advance\leftskip\mathindent
   \hscodestyle
   \let\\=\@normalcr
   \(\parray}%
  {\endparray\)%
   \hsnewpar\belowdisplayskip
   \ignorespacesafterend}

  {\parray}{\endparray}

  {\(\parray}{\endparray\)}

\def\codeframewidth{\arrayrulewidth}
\RequirePackage{calc}

  {\parskip=\abovedisplayskip\par\noindent
   \hscodestyle
   \arrayrulewidth=\codeframewidth
   \tabular{@{}|p{\linewidth-2\arraycolsep-2\arrayrulewidth-2pt}|@{}}%
   \hline\framedhslinecorrect\\{-1.5ex}%
   \let\endoflinesave=\\
   \let\\=\@normalcr
   \(\pboxed}%
  {\endpboxed\)%
   \framedhslinecorrect\endoflinesave{.5ex}\hline
   \endtabular
   \parskip=\belowdisplayskip\par\noindent
   \ignorespacesafterend}

\newcommand{\framedhslinecorrect}[2]%
  {#1[#2]}

  {\(\def\column##1##2{}%
   \let\>\undefined\let\<\undefined\let\\\undefined
   \newcommand\>[1][]{}\newcommand\<[1][]{}\newcommand\\[1][]{}%
   \def\fromto##1##2##3{##3}%
   }{\) }%

  {\let\orighscode=\hscode
   \let\origendhscode=\endhscode
   \def\endhscode{\def\hscode{\endgroup\def\@currenvir{hscode}\\}\begingroup}
   \orighscode\def\hscode{\endgroup\def\@currenvir{hscode}}}%
  {\origendhscode
   \global\let\hscode=\orighscode
   \global\let\endhscode=\origendhscode}%

\makeatother
\EndFmtInput
\ReadOnlyOnce{jfpcompat.fmt}%
\makeatletter
\def\@authortable{%
  \leavevmode \hbox \bgroup $\col@sep\tabcolsep
  \let\d@llarbegin\begingroup
  \let\d@llarend\endgroup
  \let\\\author@tabcrone
  \ignorespaces
  \@tabarray}

\makeatother
\EndFmtInput

\title{Higher-Order Recursion Abstraction: \\ How to Make Ackermann, Knuth and Conway Look Like a Bunch~of~Primitives, Figuratively Speaking}

\author{Baltasar Tranc{\'o}n y Widemann \\ Ilmenau University of Technology, DE \\ \href{mailto:baltasar.trancon@tu-ilmenau.de}{\ttfamily baltasar.trancon@tu-ilmenau.de}}

\begin{document}

\arraycolsep=0.33ex

\maketitle

\begin{abstract} 

The Ackermann function is a famous total recursive binary function on
the natural numbers.  It is the archetypal example of such a
function that is not primitive recursive, in the sense of classical
recursion theory.  However, and in seeming contradiction, there are
generalized notions of total recursion, for which the Ackermann
function is in fact primitive recursive, and often featured as a
witness for the additional power gained by the generalization.  Here,
we investigate techniques for finding and analyzing the primitive form
of complicated recursive functions, namely also Knuth's and Conway's
arrow notations, in particular by recursion abstraction, in a
framework of functional program transformation.

\end{abstract}


\section{Introduction}

The Ackermann function is a famous total binary
function on the natural numbers, given by the recursive equations:
\begin{equation*}
  \begin{array}{rcl@{\qquad}r}
    A(0, n) &=& n + 1
    \\
    A(m, 0) &=& A(m - 1, 1) & (m > 0)
    \\
    A(m, n) &=& A\bigl(m - 1, A(m, n - 1)\bigr) & (m, n > 0)
  \end{array}
\end{equation*}

It is not a particularly useful function in practice; its fame (and
ubiquitous appearance in undergraduate computer science curricula)
stems from the fact that it is one of the simplest examples of a
total, recursive (that is, recursively computable) function which is
not \emph{primitive} recursive.  The class of primitive recursive
functions is defined inductively, characteristically involving a
recursion operator that, simply speaking, allows the value of a
function at $n$ depend on its value at $n - 1$, without explicit self-reference.

Intuitively (and provably), the doubly nested self-reference of the
Ackermann function in the third equation refuses to be adapted to the
primitive recursion scheme.  But one may find seemingly contradictory
statements: the notion of primitive recursion is relative to its
underlying datatype $\mathbb{N}$; the Ackermann function \emph{does}
have a representation in terms of analogous recursion operators for
suitable, more powerful datatypes~\cite{Reynolds1985,Cockett1992}.  A
systematic introduction to recursion operators on
lists~\cite{Hutton1999} gives the Ackermann function (disguised as a
list function) as a high-end example, but does not discuss the
generalization to related functions.  Here, we start off where they
concluded.  The purpose of the present article is threefold:
\begin{enumerate} \item Recall from \cite{Hutton1999} how to obtain
a ``primitive'' representation of Ackermann's and related functions by
means of systematic \emph{recursion abstraction} in a higher-order
functional programming language.  \item Pinpoint how exactly these
representations are ``cheating'' with respect to the original,
first-order definition of primitive recursion.  \item Generalize
the approach to (apparently) even more complicated recursive
functions.  \end{enumerate}

We begin by summarizing formal details about primitive recursion in
mathematical theory and in functional programming.  The expert reader
is welcome to safely skip the next, preliminary section.  The
remainder of the article is about three concrete examples, in increasing
order of logical complexity: the Ackermann function (archetypal),
Knuth's up-arrows (slightly more powerful), Conway's chained arrows
(significantly more powerful).  They form a well-known hierarchy of
operators for the arithmetics of ``insanely large'' numbers.

Even the non-expert reader might enjoy a quick glance ahead at Figures
\ref{fig:trad} and \ref{fig:prim} on page~\pageref{fig:trad}, where
the classical and primitive definition forms are contrasted in
synopsis.  With a little contemplation, all elements of the latter,
except the \ensuremath{\Varid{fold}} operators, can be found verbatim in the former.  The
remainder of this article is a thorough exploration of the formal
details that constitute their respective equivalence.

The program code scattered throughout this article is a valid literate
Haskell program.  It makes use of standard Haskell'98 features only,
with the exception of some \texttt{Rank2Types} in meta-level analyses.
Note that, for better fit with the mathematical notation, the
first-fit rule of Haskell pattern matching is shunned; all overlapped
patterns are explicitly guarded.

\section{Primitive Recursion}

\subsection{Primitive Recursion in Mathematics}

The conventional definition of the class of primitive recursive
functions is inductive.  It contains the nullary \emph{zero} function
and the unary \emph{successor} function, and the projections, is
closed under composition and under a certain recursion operator:

Let $e$ and $g$ be $k$-ary and $(k+2)$-ary primitive recursive
functions, respectively.  Then the unique $(k+1)$-ary function $h$
given by the recursive equations
\begin{equation*}
  \begin{array}{rcl@{\qquad}r}
    h(x_1, \dots, x_k; 0) &=& e(x_1, \dots, x_k)
    \\
    h(x_1, \dots, x_k; n) &=& g\bigl(x_1, \dots, x_k; n - 1, h(x_1, \dots, x_k; n - 1)\bigr) & (n > 0)
  \end{array}
\end{equation*}%
is also primitive recursive.  Then the self-referential definition of
$h$ can be replaced equivalently by a ``primitive'' one, specifying
$e$ and $g$ instead.  It has been shown by Ackermann that a precursor
to the function now named after him is \emph{not} of this form.
Double recursion~\cite{Robinson1948} has been introduced ad-hoc for
Ackermann's and related functions, but obviously a nesting depth of
two is no natural complexity limit---why not have triple recursion, and
so forth?  Less arbitrary limits are given by type systems that allow
recursive computations on other datatypes than just natural numbers;
see below.

\subsection{Primitive Recursion in Functional Programming}

The algebraic approach to functional
programming~\cite{Meijer1991,Bird1997} generalizes the notion of a
recursion operator to arbitrary algebraic datatypes (and with some
effort, beyond).  In a functional programming context, one can do away
with the auxiliary arguments $x_1, \dots, x_k$ to the generating
functions $e$ and $g$, by having them implicitly as free variables in the
defining expression.  One can also do away with the $(k+1)$-th
argument without losing essential power~\cite{Meertens1992}, thus
arriving at a scheme called variously \emph{structural recursion},
\emph{iteration}, \emph{fold} or \emph{catamorphism}.

In the base case of the natural numbers (represented here by the
Haskell datatype \ensuremath{\Conid{Integer}}, tacitly excluding negative numbers and
\ensuremath{\bot }), the operator is:
\begin{hscode}\SaveRestoreHook
\column{B}{@{}>{\hspre}l<{\hspost}@{}}%
\column{3}{@{}>{\hspre}l<{\hspost}@{}}%
\column{10}{@{}>{\hspre}l<{\hspost}@{}}%
\column{13}{@{}>{\hspre}l<{\hspost}@{}}%
\column{16}{@{}>{\hspre}l<{\hspost}@{}}%
\column{27}{@{}>{\hspre}c<{\hspost}@{}}%
\column{27E}{@{}l@{}}%
\column{30}{@{}>{\hspre}l<{\hspost}@{}}%
\column{E}{@{}>{\hspre}l<{\hspost}@{}}%
\>[3]{}\Varid{foldn}\;{}\<[10]%
\>[10]{}\Varid{g}\;{}\<[13]%
\>[13]{}\Varid{e}\;{}\<[16]%
\>[16]{}\mathrm{0}{}\<[27]%
\>[27]{}\mathrel{=}{}\<[27E]%
\>[30]{}\Varid{e}{}\<[E]%
\\
\>[3]{}\Varid{foldn}\;{}\<[10]%
\>[10]{}\Varid{g}\;{}\<[13]%
\>[13]{}\Varid{e}\;{}\<[16]%
\>[16]{}\Varid{n}\mid \Varid{n}\mathbin{>}\mathrm{0}{}\<[27]%
\>[27]{}\mathrel{=}{}\<[27E]%
\>[30]{}\Varid{g}\;(\Varid{foldn}\;\Varid{g}\;\Varid{e}\;(\Varid{n}\mathbin{-}\mathrm{1})){}\<[E]%
\ColumnHook
\end{hscode}\resethooks
This form, as well as the constructor operations \emph{zero} and
\emph{succ}, can be deduced from the category-theoretic representation
of the natural numbers as the initial algebra of a certain functor.
We do not repeat the details here, but one consequence is highly
relevant: Fold operators have the nice universal property of
\emph{uniqueness}.  On one hand, we can of course retrieve the
preceding definition, where the generated function is named \emph{h}:

\pagebreak
From \ensuremath{\Varid{h}\equiv \Varid{foldn}\;\Varid{g}\;\Varid{e}} we can conclude
\begin{hscode}\SaveRestoreHook
\column{B}{@{}>{\hspre}l<{\hspost}@{}}%
\column{3}{@{}>{\hspre}l<{\hspost}@{}}%
\column{6}{@{}>{\hspre}l<{\hspost}@{}}%
\column{17}{@{}>{\hspre}c<{\hspost}@{}}%
\column{17E}{@{}l@{}}%
\column{21}{@{}>{\hspre}l<{\hspost}@{}}%
\column{E}{@{}>{\hspre}l<{\hspost}@{}}%
\>[3]{}\Varid{h}\;{}\<[6]%
\>[6]{}\mathrm{0}{}\<[17]%
\>[17]{}\equiv {}\<[17E]%
\>[21]{}\Varid{e}{}\<[E]%
\\
\>[3]{}\Varid{h}\;{}\<[6]%
\>[6]{}\Varid{n}\mid \Varid{n}\mathbin{>}\mathrm{0}{}\<[17]%
\>[17]{}\equiv {}\<[17E]%
\>[21]{}\Varid{g}\;(\Varid{h}\;(\Varid{n}\mathbin{-}\mathrm{1})){}\<[E]%
\ColumnHook
\end{hscode}\resethooks
But, on the other hand, the converse does also hold: From
\begin{hscode}\SaveRestoreHook
\column{B}{@{}>{\hspre}l<{\hspost}@{}}%
\column{3}{@{}>{\hspre}l<{\hspost}@{}}%
\column{6}{@{}>{\hspre}l<{\hspost}@{}}%
\column{17}{@{}>{\hspre}c<{\hspost}@{}}%
\column{17E}{@{}l@{}}%
\column{21}{@{}>{\hspre}l<{\hspost}@{}}%
\column{E}{@{}>{\hspre}l<{\hspost}@{}}%
\>[3]{}\Varid{h}\;{}\<[6]%
\>[6]{}\mathrm{0}{}\<[17]%
\>[17]{}\equiv {}\<[17E]%
\>[21]{}\Varid{e}{}\<[E]%
\\
\>[3]{}\Varid{h}\;{}\<[6]%
\>[6]{}\Varid{n}\mid \Varid{n}\mathbin{>}\mathrm{0}{}\<[17]%
\>[17]{}\equiv {}\<[17E]%
\>[21]{}\Varid{g}\;(\Varid{h}\;(\Varid{n}\mathbin{-}\mathrm{1})){}\<[E]%
\ColumnHook
\end{hscode}\resethooks
we may deduce that necessarily \ensuremath{\Varid{h}\equiv \Varid{foldn}\;\Varid{g}\;\Varid{e}}.

This will become our tool of recursion abstraction in the following
sections.  As pointed out in~\cite{Hutton1999} for the Ackermann
example, the characteristic feature of ``morally primitive'' recursive
functions is that this abstraction needs to applied more than once.

The resulting recursive functions do not look particular powerful at
first glance: mathematically, \ensuremath{\Varid{foldn}\;\Varid{g}\;\Varid{e}\;\Varid{n}} is simply $g^n(e)$.  Their
hidden power comes from the polymorphic nature of the fold operator,
which can define functions from the natural numbers to arbitrary
underlying datatypes.
\begin{hscode}\SaveRestoreHook
\column{B}{@{}>{\hspre}l<{\hspost}@{}}%
\column{3}{@{}>{\hspre}l<{\hspost}@{}}%
\column{E}{@{}>{\hspre}l<{\hspost}@{}}%
\>[3]{}\mathbf{type}\;\Conid{FoldN}\;\Varid{a}\mathrel{=}(\Varid{a}\to \Varid{a})\to \Varid{a}\to \Conid{Integer}\to \Varid{a}{}\<[E]%
\\[\blanklineskip]%
\>[3]{}\Varid{foldn}\mathbin{::}\Conid{FoldN}\;\Varid{a}{}\<[E]%
\ColumnHook
\end{hscode}\resethooks

The fold operator for lists is arguably the most popular one; see
\cite{Hutton1999}.  This is not only due to the fact that lists are
the workhorse of datatypes: List folding is both reasonably simple
regarding type and universal properties, and gives to nontrivial,
useful recursive algorithms already from very simple generator
functions.
\begin{hscode}\SaveRestoreHook
\column{B}{@{}>{\hspre}l<{\hspost}@{}}%
\column{3}{@{}>{\hspre}l<{\hspost}@{}}%
\column{23}{@{}>{\hspre}c<{\hspost}@{}}%
\column{23E}{@{}l@{}}%
\column{26}{@{}>{\hspre}l<{\hspost}@{}}%
\column{E}{@{}>{\hspre}l<{\hspost}@{}}%
\>[3]{}\Varid{foldr}\;\Varid{g}\;\Varid{e}\;[\mskip1.5mu \mskip1.5mu]{}\<[23]%
\>[23]{}\mathrel{=}{}\<[23E]%
\>[26]{}\Varid{e}{}\<[E]%
\\
\>[3]{}\Varid{foldr}\;\Varid{g}\;\Varid{e}\;(\Varid{a}\mathbin{:}\Varid{xs}){}\<[23]%
\>[23]{}\mathrel{=}{}\<[23E]%
\>[26]{}\Varid{g}\;\Varid{a}\;(\Varid{foldr}\;\Varid{g}\;\Varid{e}\;\Varid{xs}){}\<[E]%
\ColumnHook
\end{hscode}\resethooks
The type scheme for list folding is
\begin{hscode}\SaveRestoreHook
\column{B}{@{}>{\hspre}l<{\hspost}@{}}%
\column{3}{@{}>{\hspre}l<{\hspost}@{}}%
\column{E}{@{}>{\hspre}l<{\hspost}@{}}%
\>[3]{}\mathbf{type}\;\Conid{FoldR}\;\Varid{a}\;\Varid{b}\mathrel{=}(\Varid{a}\to \Varid{b}\to \Varid{b})\to \Varid{b}\to [\mskip1.5mu \Varid{a}\mskip1.5mu]\to \Varid{b}{}\<[E]%
\\[\blanklineskip]%
\>[3]{}\Varid{foldr}\mathbin{::}\Conid{FoldR}\;\Varid{a}\;\Varid{b}{}\<[E]%
\ColumnHook
\end{hscode}\resethooks
where we now have two type parameters; namely \ensuremath{\Varid{a}} for the list
elements to be consumed, and \ensuremath{\Varid{b}} for the underlying datatype in which
to perform the computation.

The associated universal property is that for a recursive function \ensuremath{\Varid{h}}
on lists, the equations
\begin{hscode}\SaveRestoreHook
\column{B}{@{}>{\hspre}l<{\hspost}@{}}%
\column{3}{@{}>{\hspre}l<{\hspost}@{}}%
\column{14}{@{}>{\hspre}c<{\hspost}@{}}%
\column{14E}{@{}l@{}}%
\column{15}{@{}>{\hspre}c<{\hspost}@{}}%
\column{15E}{@{}l@{}}%
\column{18}{@{}>{\hspre}l<{\hspost}@{}}%
\column{19}{@{}>{\hspre}l<{\hspost}@{}}%
\column{E}{@{}>{\hspre}l<{\hspost}@{}}%
\>[3]{}\Varid{h}\;[\mskip1.5mu \mskip1.5mu]{}\<[14]%
\>[14]{}\equiv {}\<[14E]%
\>[18]{}\Varid{e}{}\<[E]%
\\
\>[3]{}\Varid{h}\;(\Varid{a}\mathbin{:}\Varid{xs}){}\<[15]%
\>[15]{}\equiv {}\<[15E]%
\>[19]{}\Varid{g}\;\Varid{a}\;(\Varid{h}\;\Varid{xs}){}\<[E]%
\ColumnHook
\end{hscode}\resethooks
hold if and only if \ensuremath{\Varid{h}\equiv \Varid{foldr}\;\Varid{g}\;\Varid{e}}.

The trick behind apparently primitive representations of Ackermann and
friends is to instantiate the fold type schemes with an underlying datatype that is strictly more
powerful than the natural numbers, that is, that cannot be
G{\"o}del-encoded and -decoded by primitive recursive functions in the
original sense.

Consider a type system that has some algebraic types.  Then the class
of primitive recursive functions \emph{relative} to that type system
is defined inductively as containing some elementary functions, and
being closed under composition and the fold operators for all
algebraic datatypes, instantiated with all valid types.  

The classical definition can be retrieved as the special case of the
type system whose only datatype is $\mathbb{N}$, and whose function
types are all of the first-order form $\mathbb{N}^k \to \mathbb{N}$.
The programme of the present article is to consider functions that are not
primitive recursive in that classical sense, and to study the
derivations of their respective primitive forms in a more powerful
type system, namely one with higher-order function types, as usual in
functional programming languages.

Our chief interest lies in the program transformation techniques necessary
to take a \emph{self-referential definition}, in terms of generally
recursive equations, to a \emph{primitive definition}, that is an
expression without self-references, and composed only of legal building
blocks for primitive recursive functions.

\subsection{Elimination of Recursion}

The skeptical reader might wonder what qualifies the fold operators as
particularly primitive, seeing that their definitions make blatant use of
self-reference.  But that is an artifact of the way algebraic
datatypes are commonly defined.  An alternative definition technique,
namely \emph{Church encoding}, does not have this property.  A
Church-encoded version of the natural numbers is the following
non-recursive type definition:

\begin{hscode}\SaveRestoreHook
\column{B}{@{}>{\hspre}l<{\hspost}@{}}%
\column{3}{@{}>{\hspre}l<{\hspost}@{}}%
\column{E}{@{}>{\hspre}l<{\hspost}@{}}%
\>[3]{}\mathbf{type}\;\Conid{Nat}\mathrel{=}\forall\;\Varid{a}\mathpunct.(\Varid{a}\to \Varid{a})\to \Varid{a}\to \Varid{a}{}\<[E]%
\ColumnHook
\end{hscode}\resethooks

It allows the construction of numbers by the usual operations, by
strategic deferral of their respective interpretations \ensuremath{\Varid{z}}
and \ensuremath{\Varid{s}}:

\noindent\begin{minipage}{.5\linewidth}
\begin{hscode}\SaveRestoreHook
\column{B}{@{}>{\hspre}l<{\hspost}@{}}%
\column{3}{@{}>{\hspre}l<{\hspost}@{}}%
\column{E}{@{}>{\hspre}l<{\hspost}@{}}%
\>[3]{}\Varid{zero}\mathbin{::}\Conid{Nat}{}\<[E]%
\\
\>[3]{}\Varid{zero}\;\Varid{s}\;\Varid{z}\mathrel{=}\Varid{z}{}\<[E]%
\ColumnHook
\end{hscode}\resethooks
\end{minipage}\begin{minipage}{.5\linewidth}
\begin{hscode}\SaveRestoreHook
\column{B}{@{}>{\hspre}l<{\hspost}@{}}%
\column{3}{@{}>{\hspre}l<{\hspost}@{}}%
\column{E}{@{}>{\hspre}l<{\hspost}@{}}%
\>[3]{}\Varid{succ}\mathbin{::}\Conid{Nat}\to \Conid{Nat}{}\<[E]%
\\
\>[3]{}\Varid{succ}\;\Varid{n}\;\Varid{s}\;\Varid{z}\mathrel{=}\Varid{s}\;(\Varid{n}\;\Varid{s}\;\Varid{z}){}\<[E]%
\ColumnHook
\end{hscode}\resethooks
\end{minipage}

A trivial coercion to standard Haskell numbers can be given by
providing the obvious interpretations.
\begin{hscode}\SaveRestoreHook
\column{B}{@{}>{\hspre}l<{\hspost}@{}}%
\column{3}{@{}>{\hspre}l<{\hspost}@{}}%
\column{E}{@{}>{\hspre}l<{\hspost}@{}}%
\>[3]{}\Varid{toInteger}\mathbin{::}\Conid{Nat}\to \Conid{Integer}{}\<[E]%
\\
\>[3]{}\Varid{toInteger}\;\Varid{n}\mathrel{=}\Varid{n}\;(\mathbin{+}\mathrm{1})\;\mathrm{0}{}\<[E]%
\ColumnHook
\end{hscode}\resethooks

But, less trivially, every Church-encoded number implements the action
of the fold operator on itself, by substituting \ensuremath{\Varid{g}} for \ensuremath{\Varid{s}} and \ensuremath{\Varid{e}}
for \ensuremath{\Varid{z}}, respectively, which can thus be defined without
self-reference as
\begin{hscode}\SaveRestoreHook
\column{B}{@{}>{\hspre}l<{\hspost}@{}}%
\column{3}{@{}>{\hspre}l<{\hspost}@{}}%
\column{E}{@{}>{\hspre}l<{\hspost}@{}}%
\>[3]{}\Varid{foldn'}\mathbin{::}(\Varid{a}\to \Varid{a})\to \Varid{a}\to \Conid{Nat}\to \Varid{a}{}\<[E]%
\\
\>[3]{}\Varid{foldn'}\;\Varid{g}\;\Varid{e}\;\Varid{n}\mathrel{=}\Varid{n}\;\Varid{g}\;\Varid{e}{}\<[E]%
\ColumnHook
\end{hscode}\resethooks
and is easily seen to be equivalent to the original:
\begin{hscode}\SaveRestoreHook
\column{B}{@{}>{\hspre}l<{\hspost}@{}}%
\column{3}{@{}>{\hspre}l<{\hspost}@{}}%
\column{E}{@{}>{\hspre}l<{\hspost}@{}}%
\>[3]{}\Varid{foldn'}\;\Varid{g}\;\Varid{e}\equiv \Varid{foldn}\;\Varid{g}\;\Varid{e}\mathbin{\circ}\Varid{toInteger}{}\<[E]%
\ColumnHook
\end{hscode}\resethooks

For a general discussion of this construction for arbitrary algebraic
datatype signatures, and its equivalence to the initial algebra
approach, see \cite{Wadler1990}.

\begin{figure*}
  \fboxsep2ex
  \begin{equation*}
    \fbox{$
    \begin{array}{rcl@{\qquad}r}
      A(0, n) &=& n + 1 
      \\
      A(m, 0) &=& A(m - 1, 1) & (m > 0)
      \\
      A(m, n) &=& A(m - 1, A(m, n - 1)) & (m, n > 0)
      \\[1em]
  a \uparrow^0 b &=& a\, b
  \\         
  a \uparrow^n 0 &=& 1
  \\
  a \uparrow^n b &=& a \uparrow^{n-1} (a \uparrow^n (b - 1)) & (n, b > 0)
  \\[1em]
  \langle p \to q\rangle &=& (p+1)^{(q+1)}
  \\
  \langle X \to p \to 0\rangle &=& \langle X \to p\rangle
  \\
  \langle X \to 0 \to q\rangle &=& \langle X \to 0\rangle
  \\
  \langle X \to p \to q\rangle &=& \langle X \to (\langle X \to (p - 1) \to q\rangle - 1) \to (q - 1)\rangle
           & (p, q > 0)
\end{array}$}
\end{equation*}

  \caption{Exercise---Ackermann, Knuth and Conway in traditional appearance}
  \label{fig:trad}
\end{figure*}              

\begin{figure*}

\fbox{\begin{minipage}{0.88\textwidth}
\begin{hscode}\SaveRestoreHook
\column{B}{@{}>{\hspre}l<{\hspost}@{}}%
\column{3}{@{}>{\hspre}l<{\hspost}@{}}%
\column{5}{@{}>{\hspre}l<{\hspost}@{}}%
\column{10}{@{}>{\hspre}c<{\hspost}@{}}%
\column{10E}{@{}l@{}}%
\column{13}{@{}>{\hspre}l<{\hspost}@{}}%
\column{30}{@{}>{\hspre}l<{\hspost}@{}}%
\column{51}{@{}>{\hspre}l<{\hspost}@{}}%
\column{71}{@{}>{\hspre}l<{\hspost}@{}}%
\column{80}{@{}>{\hspre}l<{\hspost}@{}}%
\column{83}{@{}>{\hspre}l<{\hspost}@{}}%
\column{100}{@{}>{\hspre}l<{\hspost}@{}}%
\column{103}{@{}>{\hspre}l<{\hspost}@{}}%
\column{E}{@{}>{\hspre}l<{\hspost}@{}}%
\>[3]{}\Varid{ack}{}\<[10]%
\>[10]{}\mathrel{=}{}\<[10E]%
\>[30]{}\Varid{foldn}\;(\lambda \Varid{f}\to \Varid{foldn}\;{}\<[51]%
\>[51]{}\Varid{f}\;{}\<[71]%
\>[71]{}(\Varid{f}\;\mathrm{1}){}\<[80]%
\>[80]{})\;{}\<[100]%
\>[100]{}(\mathbin{+}\mathrm{1}){}\<[E]%
\\
\>[3]{}\Varid{knuth}{}\<[10]%
\>[10]{}\mathrel{=}{}\<[10E]%
\>[30]{}\Varid{foldn}\;(\lambda \Varid{f}\to \Varid{foldn}\;{}\<[51]%
\>[51]{}\Varid{f}\;{}\<[71]%
\>[71]{}\mathrm{1}{}\<[80]%
\>[80]{}){}\<[100]%
\>[100]{}\mathbin{\circ}{}\<[103]%
\>[103]{}(\mathbin{*}){}\<[E]%
\\
\>[3]{}\Varid{cback}{}\<[10]%
\>[10]{}\mathrel{=}{}\<[10E]%
\>[13]{}\Varid{foldr}\;(\lambda \Varid{o}\;\Varid{k}\to {}\<[30]%
\>[30]{}\Varid{foldn}\;(\lambda \Varid{f}\to \Varid{foldn}\;{}\<[51]%
\>[51]{}(\Varid{f}\mathbin{\circ}(-\;\mathrm{1}))\;{}\<[71]%
\>[71]{}(\Varid{k}\;\mathrm{0}\;\Varid{o}){}\<[80]%
\>[80]{})\;{}\<[83]%
\>[83]{}(\lambda \Varid{p}\to \Varid{k}\;\Varid{p}\;\Varid{o}))\;{}\<[100]%
\>[100]{}\Varid{cpow}{}\<[E]%
\\[\blanklineskip]%
\>[3]{}\hsindent{2}{}\<[5]%
\>[5]{}\mathbf{where}\;\Varid{cpow}\;\Varid{q}\;\Varid{p}\mathrel{=}(\Varid{p}\mathbin{+}\mathrm{1})\mathbin{\hat{\enspace}}(\Varid{q}\mathbin{+}\mathrm{1}){}\<[E]%
\ColumnHook
\end{hscode}\resethooks
\end{minipage}}\vspace{\belowdisplayskip}
          
  \caption{Solution---Ackermann, Knuth and Conway looking like a bunch of primitives}
  \label{fig:prim}
\end{figure*}              

\pagebreak
\section{The Ackermann Function}

\subsection{Derivation}

We start with the Haskell version of the original self-referential definition
\begin{hscode}\SaveRestoreHook
\column{B}{@{}>{\hspre}l<{\hspost}@{}}%
\column{3}{@{}>{\hspre}l<{\hspost}@{}}%
\column{9}{@{}>{\hspre}l<{\hspost}@{}}%
\column{12}{@{}>{\hspre}l<{\hspost}@{}}%
\column{32}{@{}>{\hspre}c<{\hspost}@{}}%
\column{32E}{@{}l@{}}%
\column{35}{@{}>{\hspre}l<{\hspost}@{}}%
\column{E}{@{}>{\hspre}l<{\hspost}@{}}%
\>[3]{}\Varid{ack}_0\;{}\<[9]%
\>[9]{}\mathrm{0}\;{}\<[12]%
\>[12]{}\Varid{n}{}\<[32]%
\>[32]{}\mathrel{=}{}\<[32E]%
\>[35]{}\Varid{n}\mathbin{+}\mathrm{1}{}\<[E]%
\\
\>[3]{}\Varid{ack}_0\;{}\<[9]%
\>[9]{}\Varid{m}\;{}\<[12]%
\>[12]{}\mathrm{0}\mid \Varid{m}\mathbin{>}\mathrm{0}{}\<[32]%
\>[32]{}\mathrel{=}{}\<[32E]%
\>[35]{}\Varid{ack}_0\;(\Varid{m}\mathbin{-}\mathrm{1})\;\mathrm{1}{}\<[E]%
\\
\>[3]{}\Varid{ack}_0\;{}\<[9]%
\>[9]{}\Varid{m}\;{}\<[12]%
\>[12]{}\Varid{n}\mid \Varid{m}\mathbin{>}\mathrm{0}\mathrel{\wedge}\Varid{n}\mathbin{>}\mathrm{0}{}\<[32]%
\>[32]{}\mathrel{=}{}\<[32E]%
\>[35]{}\Varid{ack}_0\;(\Varid{m}\mathbin{-}\mathrm{1})\;(\Varid{ack}_0\;\Varid{m}\;(\Varid{n}\mathbin{-}\mathrm{1})){}\<[E]%
\ColumnHook
\end{hscode}\resethooks
and proceed by successive application of elementary transformation
steps.  Each new version is indexed differently, such that all can
coexist in a single literate Haskell source file for this whole article.

In a first transformation step, we exploit currying to eliminate the
second function argument \ensuremath{\Varid{n}} and thus unify the second and third
equation in an auxiliary function \ensuremath{\Varid{aux}}.
\begin{hscode}\SaveRestoreHook
\column{B}{@{}>{\hspre}l<{\hspost}@{}}%
\column{3}{@{}>{\hspre}l<{\hspost}@{}}%
\column{5}{@{}>{\hspre}l<{\hspost}@{}}%
\column{9}{@{}>{\hspre}l<{\hspost}@{}}%
\column{13}{@{}>{\hspre}l<{\hspost}@{}}%
\column{18}{@{}>{\hspre}l<{\hspost}@{}}%
\column{20}{@{}>{\hspre}c<{\hspost}@{}}%
\column{20E}{@{}l@{}}%
\column{23}{@{}>{\hspre}l<{\hspost}@{}}%
\column{29}{@{}>{\hspre}c<{\hspost}@{}}%
\column{29E}{@{}l@{}}%
\column{32}{@{}>{\hspre}l<{\hspost}@{}}%
\column{E}{@{}>{\hspre}l<{\hspost}@{}}%
\>[3]{}\Varid{ack}_1\;{}\<[9]%
\>[9]{}\mathrm{0}{}\<[20]%
\>[20]{}\mathrel{=}{}\<[20E]%
\>[23]{}(\mathbin{+}\mathrm{1}){}\<[E]%
\\
\>[3]{}\Varid{ack}_1\;{}\<[9]%
\>[9]{}\Varid{m}\mid \Varid{m}\mathbin{>}\mathrm{0}{}\<[20]%
\>[20]{}\mathrel{=}{}\<[20E]%
\>[23]{}\Varid{aux}{}\<[E]%
\\
\>[3]{}\hsindent{2}{}\<[5]%
\>[5]{}\mathbf{where}\;{}\<[13]%
\>[13]{}\Varid{aux}\;{}\<[18]%
\>[18]{}\mathrm{0}{}\<[29]%
\>[29]{}\mathrel{=}{}\<[29E]%
\>[32]{}\Varid{ack}_1\;(\Varid{m}\mathbin{-}\mathrm{1})\;\mathrm{1}{}\<[E]%
\\
\>[13]{}\Varid{aux}\;{}\<[18]%
\>[18]{}\Varid{n}\mid \Varid{n}\mathbin{>}\mathrm{0}{}\<[29]%
\>[29]{}\mathrel{=}{}\<[29E]%
\>[32]{}\Varid{ack}_1\;(\Varid{m}\mathbin{-}\mathrm{1})\;(\Varid{ack}_{1\mathrm{a}}\;\Varid{m}\;(\Varid{n}\mathbin{-}\mathrm{1})){}\<[E]%
\ColumnHook
\end{hscode}\resethooks

We factor out the common term \ensuremath{\Varid{ack}\;(\Varid{m}\mathbin{-}\mathrm{1})} by beta expansion to an
argument \ensuremath{\Varid{f}}.  What seems like a simple matter of abbreviation here
will prove a very useful preparation step later.
\begin{hscode}\SaveRestoreHook
\column{B}{@{}>{\hspre}l<{\hspost}@{}}%
\column{3}{@{}>{\hspre}l<{\hspost}@{}}%
\column{5}{@{}>{\hspre}l<{\hspost}@{}}%
\column{10}{@{}>{\hspre}l<{\hspost}@{}}%
\column{12}{@{}>{\hspre}l<{\hspost}@{}}%
\column{17}{@{}>{\hspre}l<{\hspost}@{}}%
\column{20}{@{}>{\hspre}l<{\hspost}@{}}%
\column{21}{@{}>{\hspre}c<{\hspost}@{}}%
\column{21E}{@{}l@{}}%
\column{24}{@{}>{\hspre}l<{\hspost}@{}}%
\column{31}{@{}>{\hspre}c<{\hspost}@{}}%
\column{31E}{@{}l@{}}%
\column{34}{@{}>{\hspre}l<{\hspost}@{}}%
\column{E}{@{}>{\hspre}l<{\hspost}@{}}%
\>[3]{}\Varid{ack}_{1\mathrm{a}}\;{}\<[10]%
\>[10]{}\mathrm{0}{}\<[21]%
\>[21]{}\mathrel{=}{}\<[21E]%
\>[24]{}(\mathbin{+}\mathrm{1}){}\<[E]%
\\
\>[3]{}\Varid{ack}_{1\mathrm{a}}\;{}\<[10]%
\>[10]{}\Varid{m}\mid \Varid{m}\mathbin{>}\mathrm{0}{}\<[21]%
\>[21]{}\mathrel{=}{}\<[21E]%
\>[24]{}\Varid{aux}\;(\Varid{ack}_{1\mathrm{a}}\;(\Varid{m}\mathbin{-}\mathrm{1})){}\<[E]%
\\
\>[3]{}\hsindent{2}{}\<[5]%
\>[5]{}\mathbf{where}\;{}\<[12]%
\>[12]{}\Varid{aux}\;{}\<[17]%
\>[17]{}\Varid{f}\;{}\<[20]%
\>[20]{}\mathrm{0}{}\<[31]%
\>[31]{}\mathrel{=}{}\<[31E]%
\>[34]{}\Varid{f}\;\mathrm{1}{}\<[E]%
\\
\>[12]{}\Varid{aux}\;{}\<[17]%
\>[17]{}\Varid{f}\;{}\<[20]%
\>[20]{}\Varid{n}\mid \Varid{n}\mathbin{>}\mathrm{0}{}\<[31]%
\>[31]{}\mathrel{=}{}\<[31E]%
\>[34]{}\Varid{f}\;(\Varid{ack}_{1\mathrm{a}}\;\Varid{m}\;(\Varid{n}\mathbin{-}\mathrm{1})){}\<[E]%
\ColumnHook
\end{hscode}\resethooks

The single most tricky point is to get rid of the back-reference from
\ensuremath{\Varid{aux}} to \ensuremath{\Varid{ack}}.  Note that \ensuremath{\Varid{aux}} is only ever applied to \ensuremath{\Varid{f}\mathrel{=}\Varid{ack}\;(\Varid{m}\mathbin{-}\mathrm{1})}, hence we may substitute\footnote{In case of doubt about the
validity of this slightly simplified argument, see
section~\ref{conclusion} for details.}
\begin{hscode}\SaveRestoreHook
\column{B}{@{}>{\hspre}l<{\hspost}@{}}%
\column{3}{@{}>{\hspre}l<{\hspost}@{}}%
\column{E}{@{}>{\hspre}l<{\hspost}@{}}%
\>[3]{}\Varid{ack}\;\Varid{m}\equiv \Varid{aux}\;(\Varid{ack}\;(\Varid{m}\mathbin{-}\mathrm{1}))\equiv \Varid{aux}\;\Varid{f}{}\<[E]%
\ColumnHook
\end{hscode}\resethooks
and obtain:
\begin{hscode}\SaveRestoreHook
\column{B}{@{}>{\hspre}l<{\hspost}@{}}%
\column{3}{@{}>{\hspre}l<{\hspost}@{}}%
\column{5}{@{}>{\hspre}l<{\hspost}@{}}%
\column{10}{@{}>{\hspre}l<{\hspost}@{}}%
\column{12}{@{}>{\hspre}l<{\hspost}@{}}%
\column{17}{@{}>{\hspre}l<{\hspost}@{}}%
\column{20}{@{}>{\hspre}l<{\hspost}@{}}%
\column{21}{@{}>{\hspre}c<{\hspost}@{}}%
\column{21E}{@{}l@{}}%
\column{24}{@{}>{\hspre}l<{\hspost}@{}}%
\column{31}{@{}>{\hspre}c<{\hspost}@{}}%
\column{31E}{@{}l@{}}%
\column{34}{@{}>{\hspre}l<{\hspost}@{}}%
\column{E}{@{}>{\hspre}l<{\hspost}@{}}%
\>[3]{}\Varid{ack}_{1\mathrm{b}}\;{}\<[10]%
\>[10]{}\mathrm{0}{}\<[21]%
\>[21]{}\mathrel{=}{}\<[21E]%
\>[24]{}(\mathbin{+}\mathrm{1}){}\<[E]%
\\
\>[3]{}\Varid{ack}_{1\mathrm{b}}\;{}\<[10]%
\>[10]{}\Varid{m}\mid \Varid{m}\mathbin{>}\mathrm{0}{}\<[21]%
\>[21]{}\mathrel{=}{}\<[21E]%
\>[24]{}\Varid{aux}\;(\Varid{ack}_{1\mathrm{b}}\;(\Varid{m}\mathbin{-}\mathrm{1})){}\<[E]%
\\
\>[3]{}\hsindent{2}{}\<[5]%
\>[5]{}\mathbf{where}\;{}\<[12]%
\>[12]{}\Varid{aux}\;{}\<[17]%
\>[17]{}\Varid{f}\;{}\<[20]%
\>[20]{}\mathrm{0}{}\<[31]%
\>[31]{}\mathrel{=}{}\<[31E]%
\>[34]{}\Varid{f}\;\mathrm{1}{}\<[E]%
\\
\>[12]{}\Varid{aux}\;{}\<[17]%
\>[17]{}\Varid{f}\;{}\<[20]%
\>[20]{}\Varid{n}\mid \Varid{n}\mathbin{>}\mathrm{0}{}\<[31]%
\>[31]{}\mathrel{=}{}\<[31E]%
\>[34]{}\Varid{f}\;(\Varid{aux}\;\Varid{f}\;(\Varid{n}\mathbin{-}\mathrm{1})){}\<[E]%
\ColumnHook
\end{hscode}\resethooks

Now we have \ensuremath{\Varid{ack}} in a form generated by the fold operator from a
function \ensuremath{\Varid{aux}} whose definition is independent from \ensuremath{\Varid{ack}}.  We may invoke the
universal property and conclude:
\begin{hscode}\SaveRestoreHook
\column{B}{@{}>{\hspre}l<{\hspost}@{}}%
\column{3}{@{}>{\hspre}l<{\hspost}@{}}%
\column{5}{@{}>{\hspre}l<{\hspost}@{}}%
\column{12}{@{}>{\hspre}l<{\hspost}@{}}%
\column{17}{@{}>{\hspre}l<{\hspost}@{}}%
\column{20}{@{}>{\hspre}l<{\hspost}@{}}%
\column{31}{@{}>{\hspre}c<{\hspost}@{}}%
\column{31E}{@{}l@{}}%
\column{34}{@{}>{\hspre}l<{\hspost}@{}}%
\column{37}{@{}>{\hspre}l<{\hspost}@{}}%
\column{E}{@{}>{\hspre}l<{\hspost}@{}}%
\>[3]{}\Varid{ack}_2\mathrel{=}\Varid{foldn}\;\Varid{aux}\;(\mathbin{+}\mathrm{1}){}\<[E]%
\\
\>[3]{}\hsindent{2}{}\<[5]%
\>[5]{}\mathbf{where}\;{}\<[12]%
\>[12]{}\Varid{aux}\;{}\<[17]%
\>[17]{}\Varid{f}\;{}\<[20]%
\>[20]{}\mathrm{0}{}\<[31]%
\>[31]{}\mathrel{=}{}\<[31E]%
\>[34]{}\Varid{f}\;{}\<[37]%
\>[37]{}\mathrm{1}{}\<[E]%
\\
\>[12]{}\Varid{aux}\;{}\<[17]%
\>[17]{}\Varid{f}\;{}\<[20]%
\>[20]{}\Varid{n}\mid \Varid{n}\mathbin{>}\mathrm{0}{}\<[31]%
\>[31]{}\mathrel{=}{}\<[31E]%
\>[34]{}\Varid{f}\;{}\<[37]%
\>[37]{}(\Varid{aux}\;\Varid{f}\;(\Varid{n}\mathbin{-}\mathrm{1})){}\<[E]%
\ColumnHook
\end{hscode}\resethooks

It remains to be established that the self-referential definition of
the generator \ensuremath{\Varid{aux}} can be reduced to a primitive form.  As it
happens, it is already of the right shape.
\begin{hscode}\SaveRestoreHook
\column{B}{@{}>{\hspre}l<{\hspost}@{}}%
\column{3}{@{}>{\hspre}l<{\hspost}@{}}%
\column{5}{@{}>{\hspre}l<{\hspost}@{}}%
\column{12}{@{}>{\hspre}l<{\hspost}@{}}%
\column{E}{@{}>{\hspre}l<{\hspost}@{}}%
\>[3]{}\Varid{ack}_3\mathrel{=}\Varid{foldn}\;\Varid{aux}\;(\mathbin{+}\mathrm{1}){}\<[E]%
\\
\>[3]{}\hsindent{2}{}\<[5]%
\>[5]{}\mathbf{where}\;{}\<[12]%
\>[12]{}\Varid{aux}\;\Varid{f}\mathrel{=}\Varid{foldn}\;\Varid{f}\;(\Varid{f}\;\mathrm{1}){}\<[E]%
\ColumnHook
\end{hscode}\resethooks

So, in summary, we have the Ackermann function given by an entirely
non-self-referen\-tial expression that invokes two nested instances of
the fold operator.
\begin{hscode}\SaveRestoreHook
\column{B}{@{}>{\hspre}l<{\hspost}@{}}%
\column{3}{@{}>{\hspre}l<{\hspost}@{}}%
\column{E}{@{}>{\hspre}l<{\hspost}@{}}%
\>[3]{}\Varid{ack}_4\mathrel{=}\Varid{foldn}\;(\lambda \Varid{f}\to \Varid{foldn}\;\Varid{f}\;(\Varid{f}\;\mathrm{1}))\;(\mathbin{+}\mathrm{1}){}\<[E]%
\ColumnHook
\end{hscode}\resethooks

\subsection{Verification}

Since the final version of the Ackermann function admittedly looks
very different from the original one, it is instructive to expand it
back and demonstrate the inductive equivalence to the original
recursive equations.
As the base case for \ensuremath{\Varid{n}} we have
\begin{hscode}\SaveRestoreHook
\column{B}{@{}>{\hspre}l<{\hspost}@{}}%
\column{3}{@{}>{\hspre}l<{\hspost}@{}}%
\column{13}{@{}>{\hspre}c<{\hspost}@{}}%
\column{13E}{@{}l@{}}%
\column{17}{@{}>{\hspre}l<{\hspost}@{}}%
\column{E}{@{}>{\hspre}l<{\hspost}@{}}%
\>[3]{}\Varid{ack}_4\;\mathrm{0}\;\Varid{n}{}\<[13]%
\>[13]{}\equiv {}\<[13E]%
\>[17]{}\Varid{foldn}\;(\lambda \Varid{f}\to \Varid{foldn}\;\Varid{f}\;(\Varid{f}\;\mathrm{1}))\;(\mathbin{+}\mathrm{1})\;\mathrm{0}\;\Varid{n}{}\<[E]%
\\
\>[13]{}\equiv {}\<[13E]%
\>[17]{}(\mathbin{+}\mathrm{1})\;\Varid{n}{}\<[E]%
\\
\>[13]{}\equiv {}\<[13E]%
\>[17]{}\Varid{n}\mathbin{+}\mathrm{1}{}\<[E]%
\ColumnHook
\end{hscode}\resethooks
and as the inductive case we have
\begin{hscode}\SaveRestoreHook
\column{B}{@{}>{\hspre}l<{\hspost}@{}}%
\column{3}{@{}>{\hspre}l<{\hspost}@{}}%
\column{19}{@{}>{\hspre}c<{\hspost}@{}}%
\column{19E}{@{}l@{}}%
\column{23}{@{}>{\hspre}l<{\hspost}@{}}%
\column{49}{@{}>{\hspre}l<{\hspost}@{}}%
\column{85}{@{}>{\hspre}c<{\hspost}@{}}%
\column{85E}{@{}l@{}}%
\column{88}{@{}>{\hspre}l<{\hspost}@{}}%
\column{E}{@{}>{\hspre}l<{\hspost}@{}}%
\>[3]{}\Varid{ack}_4\;\Varid{m}\mid \Varid{m}\mathbin{>}\mathrm{0}{}\<[19]%
\>[19]{}\equiv {}\<[19E]%
\>[49]{}\Varid{foldn}\;(\lambda \Varid{f}\to \Varid{foldn}\;\Varid{f}\;(\Varid{f}\;\mathrm{1}))\;(\mathbin{+}\mathrm{1})\;{}\<[88]%
\>[88]{}\Varid{m}{}\<[E]%
\\
\>[19]{}\equiv {}\<[19E]%
\>[23]{}(\lambda \Varid{f}\to \Varid{foldn}\;\Varid{f}\;(\Varid{f}\;\mathrm{1}))\;({}\<[49]%
\>[49]{}\Varid{foldn}\;(\lambda \Varid{f}\to \Varid{foldn}\;\Varid{f}\;(\Varid{f}\;\mathrm{1}))\;(\mathbin{+}\mathrm{1})\;{}\<[85]%
\>[85]{}({}\<[85E]%
\>[88]{}\Varid{m}\mathbin{-}\mathrm{1})){}\<[E]%
\\
\>[19]{}\equiv {}\<[19E]%
\>[23]{}(\lambda \Varid{f}\to \Varid{foldn}\;\Varid{f}\;(\Varid{f}\;\mathrm{1}))\;(\Varid{ack}_4\;{}\<[85]%
\>[85]{}({}\<[85E]%
\>[88]{}\Varid{m}\mathbin{-}\mathrm{1})){}\<[E]%
\\
\>[19]{}\equiv {}\<[19E]%
\>[23]{}\Varid{foldn}\;(\Varid{ack}_4\;(\Varid{m}\mathbin{-}\mathrm{1}))\;(\Varid{ack}_4\;(\Varid{m}\mathbin{-}\mathrm{1})\;\mathrm{1}){}\<[E]%
\ColumnHook
\end{hscode}\resethooks
which cannot be unfolded yet.  Proceeding to the second argument,
we have the base case
\begin{hscode}\SaveRestoreHook
\column{B}{@{}>{\hspre}l<{\hspost}@{}}%
\column{3}{@{}>{\hspre}l<{\hspost}@{}}%
\column{13}{@{}>{\hspre}c<{\hspost}@{}}%
\column{13E}{@{}l@{}}%
\column{17}{@{}>{\hspre}l<{\hspost}@{}}%
\column{41}{@{}>{\hspre}l<{\hspost}@{}}%
\column{E}{@{}>{\hspre}l<{\hspost}@{}}%
\>[3]{}\Varid{ack}_4\;\Varid{m}\;\mathrm{0}{}\<[13]%
\>[13]{}\equiv {}\<[13E]%
\>[17]{}\Varid{foldn}\;(\Varid{ack}_4\;(\Varid{m}\mathbin{-}\mathrm{1}))\;({}\<[41]%
\>[41]{}\Varid{ack}_4\;(\Varid{m}\mathbin{-}\mathrm{1})\;\mathrm{1})\;\mathrm{0}{}\<[E]%
\\
\>[13]{}\equiv {}\<[13E]%
\>[41]{}\Varid{ack}_4\;(\Varid{m}\mathbin{-}\mathrm{1})\;\mathrm{1}{}\<[E]%
\ColumnHook
\end{hscode}\resethooks
and the inductive case
\begin{hscode}\SaveRestoreHook
\column{B}{@{}>{\hspre}l<{\hspost}@{}}%
\column{3}{@{}>{\hspre}l<{\hspost}@{}}%
\column{21}{@{}>{\hspre}c<{\hspost}@{}}%
\column{21E}{@{}l@{}}%
\column{25}{@{}>{\hspre}l<{\hspost}@{}}%
\column{41}{@{}>{\hspre}l<{\hspost}@{}}%
\column{80}{@{}>{\hspre}c<{\hspost}@{}}%
\column{80E}{@{}l@{}}%
\column{83}{@{}>{\hspre}l<{\hspost}@{}}%
\column{E}{@{}>{\hspre}l<{\hspost}@{}}%
\>[3]{}\Varid{ack}_4\;\Varid{m}\;\Varid{n}\mid \Varid{n}\mathbin{>}\mathrm{0}{}\<[21]%
\>[21]{}\equiv {}\<[21E]%
\>[41]{}\Varid{foldn}\;(\Varid{ack}_4\;(\Varid{m}\mathbin{-}\mathrm{1}))\;(\Varid{ack}_4\;(\Varid{m}\mathbin{-}\mathrm{1})\;\mathrm{1})\;{}\<[83]%
\>[83]{}\Varid{n}{}\<[E]%
\\
\>[21]{}\equiv {}\<[21E]%
\>[25]{}\Varid{ack}_4\;(\Varid{m}\mathbin{-}\mathrm{1})\;({}\<[41]%
\>[41]{}\Varid{foldn}\;(\Varid{ack}_4\;(\Varid{m}\mathbin{-}\mathrm{1}))\;(\Varid{ack}_4\;(\Varid{m}\mathbin{-}\mathrm{1})\;\mathrm{1})\;{}\<[80]%
\>[80]{}({}\<[80E]%
\>[83]{}\Varid{n}\mathbin{-}\mathrm{1})){}\<[E]%
\\
\>[21]{}\equiv {}\<[21E]%
\>[25]{}\Varid{ack}_4\;(\Varid{m}\mathbin{-}\mathrm{1})\;({}\<[41]%
\>[41]{}\Varid{ack}_4\;\Varid{m}\;{}\<[80]%
\>[80]{}({}\<[80E]%
\>[83]{}\Varid{n}\mathbin{-}\mathrm{1})){}\<[E]%
\ColumnHook
\end{hscode}\resethooks
by back-substitution. \quad$\Box$

\subsection{Analysis}

So, in a sense, the Ackermann function is primitive after all!  As
discussed above, the polymorphic fold operator glosses over the fact
that underlying datatypes of different, incommensurable power are used
here.  To make things more explicit, consider monomorphic instances of
\ensuremath{\Varid{foldn}}:
\begin{hscode}\SaveRestoreHook
\column{B}{@{}>{\hspre}l<{\hspost}@{}}%
\column{3}{@{}>{\hspre}l<{\hspost}@{}}%
\column{E}{@{}>{\hspre}l<{\hspost}@{}}%
\>[3]{}\Varid{foldn}_1\mathbin{::}\Conid{FoldN}\;\Conid{Integer}{}\<[E]%
\\
\>[3]{}\Varid{foldn}_1\mathrel{=}\Varid{foldn}{}\<[E]%
\\[\blanklineskip]%
\>[3]{}\Varid{foldn}_2\mathbin{::}\Conid{FoldN}\;(\Conid{Integer}\to \Conid{Integer}){}\<[E]%
\\
\>[3]{}\Varid{foldn}_2\mathrel{=}\Varid{foldn}{}\<[E]%
\ColumnHook
\end{hscode}\resethooks
Using these, we can make the different roles explicit:
\begin{hscode}\SaveRestoreHook
\column{B}{@{}>{\hspre}l<{\hspost}@{}}%
\column{3}{@{}>{\hspre}l<{\hspost}@{}}%
\column{E}{@{}>{\hspre}l<{\hspost}@{}}%
\>[3]{}\Varid{ack}_5\mathrel{=}\Varid{foldn}_2\;(\lambda \Varid{f}\to \Varid{foldn}_1\;\Varid{f}\;(\Varid{f}\;\mathrm{1}))\;(\mathbin{+}\mathrm{1}){}\<[E]%
\ColumnHook
\end{hscode}\resethooks
Contradiction is avoided by the fact that the higher-order datatype
underlying the outer instance cannot be converted to and from the
traditional first-order datatype underlying the inner one by means of
primitive recursive conversion functions.

\section{Knuth's Up-Arrows}

A natural next candidate function to investigate is Knuth's up-arrow,
which is known to have quite similar but mildly more flexible behavior
than the Ackermann function.  The usual recursive definition is, for $a, b \geq 0; n > 0$:
\begin{equation*}
  \begin{array}{rcl@{\qquad}r}
  a \uparrow^1 b &=& a^b
  \\         
  a \uparrow^n 0 &=& 1
  \\
  a \uparrow^n b &=& a \uparrow^{n-1} \bigl(a \uparrow^n (b - 1)\bigr) & (n > 1; b > 0)
  \end{array}
\end{equation*}
This can be canonically extended to $n = 0$:
\begin{equation*}
  \begin{array}{rcl@{\qquad}r}
  a \uparrow^0 b &=& a\, b
  \\         
  a \uparrow^n 0 &=& 1 & (n > 0)
  \\
  a \uparrow^n b &=& a \uparrow^{n-1} \bigl(a \uparrow^n (b - 1)\bigr) & (n,b > 0)
  \end{array}
\end{equation*}%
which is also a neat case of consistent abuse of notation, seeing that 
\begin{equation*}
  a \uparrow^n b = a \underbrace{\uparrow\cdots\uparrow}_n b
\end{equation*}

\subsection{Derivation}

This definition translates straightforwardly to Haskell:
\begin{hscode}\SaveRestoreHook
\column{B}{@{}>{\hspre}l<{\hspost}@{}}%
\column{3}{@{}>{\hspre}l<{\hspost}@{}}%
\column{11}{@{}>{\hspre}l<{\hspost}@{}}%
\column{14}{@{}>{\hspre}l<{\hspost}@{}}%
\column{17}{@{}>{\hspre}l<{\hspost}@{}}%
\column{38}{@{}>{\hspre}c<{\hspost}@{}}%
\column{38E}{@{}l@{}}%
\column{41}{@{}>{\hspre}l<{\hspost}@{}}%
\column{E}{@{}>{\hspre}l<{\hspost}@{}}%
\>[3]{}\Varid{knuth}_0\;{}\<[11]%
\>[11]{}\Varid{a}\;{}\<[14]%
\>[14]{}\mathrm{0}\;{}\<[17]%
\>[17]{}\Varid{b}{}\<[38]%
\>[38]{}\mathrel{=}{}\<[38E]%
\>[41]{}\Varid{a}\mathbin{*}\Varid{b}{}\<[E]%
\\
\>[3]{}\Varid{knuth}_0\;{}\<[11]%
\>[11]{}\Varid{a}\;{}\<[14]%
\>[14]{}\Varid{n}\;{}\<[17]%
\>[17]{}\mathrm{0}\mid \Varid{n}\mathbin{>}\mathrm{0}{}\<[38]%
\>[38]{}\mathrel{=}{}\<[38E]%
\>[41]{}\mathrm{1}{}\<[E]%
\\
\>[3]{}\Varid{knuth}_0\;{}\<[11]%
\>[11]{}\Varid{a}\;{}\<[14]%
\>[14]{}\Varid{n}\;{}\<[17]%
\>[17]{}\Varid{b}\mid \Varid{n}\mathbin{>}\mathrm{0}\mathrel{\wedge}\Varid{b}\mathbin{>}\mathrm{0}{}\<[38]%
\>[38]{}\mathrel{=}{}\<[38E]%
\>[41]{}\Varid{knuth}_0\;\Varid{a}\;(\Varid{n}\mathbin{-}\mathrm{1})\;(\Varid{knuth}_0\;\Varid{a}\;\Varid{n}\;(\Varid{b}\mathbin{-}\mathrm{1})){}\<[E]%
\ColumnHook
\end{hscode}\resethooks
In analogy to the Ackermann example, we curry argument \ensuremath{\Varid{b}}.
\begin{hscode}\SaveRestoreHook
\column{B}{@{}>{\hspre}l<{\hspost}@{}}%
\column{3}{@{}>{\hspre}l<{\hspost}@{}}%
\column{5}{@{}>{\hspre}l<{\hspost}@{}}%
\column{11}{@{}>{\hspre}l<{\hspost}@{}}%
\column{12}{@{}>{\hspre}l<{\hspost}@{}}%
\column{14}{@{}>{\hspre}l<{\hspost}@{}}%
\column{17}{@{}>{\hspre}l<{\hspost}@{}}%
\column{25}{@{}>{\hspre}l<{\hspost}@{}}%
\column{28}{@{}>{\hspre}c<{\hspost}@{}}%
\column{28E}{@{}l@{}}%
\column{31}{@{}>{\hspre}l<{\hspost}@{}}%
\column{E}{@{}>{\hspre}l<{\hspost}@{}}%
\>[3]{}\Varid{knuth}_1\;{}\<[11]%
\>[11]{}\Varid{a}\;{}\<[14]%
\>[14]{}\mathrm{0}{}\<[25]%
\>[25]{}\mathrel{=}(\Varid{a}\mathbin{*}){}\<[E]%
\\
\>[3]{}\Varid{knuth}_1\;{}\<[11]%
\>[11]{}\Varid{a}\;{}\<[14]%
\>[14]{}\Varid{n}\mid \Varid{n}\mathbin{>}\mathrm{0}{}\<[25]%
\>[25]{}\mathrel{=}\Varid{aux}{}\<[E]%
\\
\>[3]{}\hsindent{2}{}\<[5]%
\>[5]{}\mathbf{where}\;{}\<[12]%
\>[12]{}\Varid{aux}\;{}\<[17]%
\>[17]{}\mathrm{0}{}\<[28]%
\>[28]{}\mathrel{=}{}\<[28E]%
\>[31]{}\mathrm{1}{}\<[E]%
\\
\>[12]{}\Varid{aux}\;{}\<[17]%
\>[17]{}\Varid{b}\mid \Varid{b}\mathbin{>}\mathrm{0}{}\<[28]%
\>[28]{}\mathrel{=}{}\<[28E]%
\>[31]{}\Varid{knuth}_1\;\Varid{a}\;(\Varid{n}\mathbin{-}\mathrm{1})\;(\Varid{knuth}_1\;\Varid{a}\;\Varid{n}\;(\Varid{b}\mathbin{-}\mathrm{1})){}\<[E]%
\ColumnHook
\end{hscode}\resethooks

If we had tried this example first, we would probably have gotten
stuck at this point!  But following the procedure for the Ackermann
function, we beta-expand the recursive subterm \ensuremath{\Varid{knuth}\;\Varid{a}\;(\Varid{n}\mathbin{-}\mathrm{1})}, even
if there is only one occurrence, and the step does not simplify
anything at first glance.
\begin{hscode}\SaveRestoreHook
\column{B}{@{}>{\hspre}l<{\hspost}@{}}%
\column{3}{@{}>{\hspre}l<{\hspost}@{}}%
\column{5}{@{}>{\hspre}l<{\hspost}@{}}%
\column{12}{@{}>{\hspre}l<{\hspost}@{}}%
\column{15}{@{}>{\hspre}l<{\hspost}@{}}%
\column{17}{@{}>{\hspre}l<{\hspost}@{}}%
\column{20}{@{}>{\hspre}l<{\hspost}@{}}%
\column{26}{@{}>{\hspre}c<{\hspost}@{}}%
\column{26E}{@{}l@{}}%
\column{29}{@{}>{\hspre}l<{\hspost}@{}}%
\column{31}{@{}>{\hspre}c<{\hspost}@{}}%
\column{31E}{@{}l@{}}%
\column{34}{@{}>{\hspre}l<{\hspost}@{}}%
\column{E}{@{}>{\hspre}l<{\hspost}@{}}%
\>[3]{}\Varid{knuth}_{1\mathrm{a}}\;{}\<[12]%
\>[12]{}\Varid{a}\;{}\<[15]%
\>[15]{}\mathrm{0}{}\<[26]%
\>[26]{}\mathrel{=}{}\<[26E]%
\>[29]{}(\Varid{a}\mathbin{*}){}\<[E]%
\\
\>[3]{}\Varid{knuth}_{1\mathrm{a}}\;{}\<[12]%
\>[12]{}\Varid{a}\;{}\<[15]%
\>[15]{}\Varid{n}\mid \Varid{n}\mathbin{>}\mathrm{0}{}\<[26]%
\>[26]{}\mathrel{=}{}\<[26E]%
\>[29]{}\Varid{aux}\;(\Varid{knuth}_{1\mathrm{a}}\;\Varid{a}\;(\Varid{n}\mathbin{-}\mathrm{1})){}\<[E]%
\\
\>[3]{}\hsindent{2}{}\<[5]%
\>[5]{}\mathbf{where}\;{}\<[12]%
\>[12]{}\Varid{aux}\;{}\<[17]%
\>[17]{}\Varid{f}\;{}\<[20]%
\>[20]{}\mathrm{0}{}\<[31]%
\>[31]{}\mathrel{=}{}\<[31E]%
\>[34]{}\mathrm{1}{}\<[E]%
\\
\>[12]{}\Varid{aux}\;{}\<[17]%
\>[17]{}\Varid{f}\;{}\<[20]%
\>[20]{}\Varid{b}\mid \Varid{b}\mathbin{>}\mathrm{0}{}\<[31]%
\>[31]{}\mathrel{=}{}\<[31E]%
\>[34]{}\Varid{f}\;(\Varid{knuth}_{1\mathrm{a}}\;\Varid{a}\;\Varid{n}\;(\Varid{b}\mathbin{-}\mathrm{1})){}\<[E]%
\ColumnHook
\end{hscode}\resethooks
We find that, in the context of the last equation, we can again substitute
\begin{hscode}\SaveRestoreHook
\column{B}{@{}>{\hspre}l<{\hspost}@{}}%
\column{3}{@{}>{\hspre}l<{\hspost}@{}}%
\column{E}{@{}>{\hspre}l<{\hspost}@{}}%
\>[3]{}\Varid{knuth}\;\Varid{a}\;\Varid{n}\equiv \Varid{aux}\;(\Varid{knuth}\;\Varid{a}\;(\Varid{n}\mathbin{-}\mathrm{1}))\equiv \Varid{aux}\;\Varid{f}{}\<[E]%
\ColumnHook
\end{hscode}\resethooks
and arrive at:
\begin{hscode}\SaveRestoreHook
\column{B}{@{}>{\hspre}l<{\hspost}@{}}%
\column{3}{@{}>{\hspre}l<{\hspost}@{}}%
\column{5}{@{}>{\hspre}l<{\hspost}@{}}%
\column{12}{@{}>{\hspre}l<{\hspost}@{}}%
\column{15}{@{}>{\hspre}l<{\hspost}@{}}%
\column{17}{@{}>{\hspre}l<{\hspost}@{}}%
\column{20}{@{}>{\hspre}l<{\hspost}@{}}%
\column{26}{@{}>{\hspre}c<{\hspost}@{}}%
\column{26E}{@{}l@{}}%
\column{29}{@{}>{\hspre}l<{\hspost}@{}}%
\column{31}{@{}>{\hspre}c<{\hspost}@{}}%
\column{31E}{@{}l@{}}%
\column{34}{@{}>{\hspre}l<{\hspost}@{}}%
\column{E}{@{}>{\hspre}l<{\hspost}@{}}%
\>[3]{}\Varid{knuth}_{1\mathrm{b}}\;{}\<[12]%
\>[12]{}\Varid{a}\;{}\<[15]%
\>[15]{}\mathrm{0}{}\<[26]%
\>[26]{}\mathrel{=}{}\<[26E]%
\>[29]{}(\Varid{a}\mathbin{*}){}\<[E]%
\\
\>[3]{}\Varid{knuth}_{1\mathrm{b}}\;{}\<[12]%
\>[12]{}\Varid{a}\;{}\<[15]%
\>[15]{}\Varid{n}\mid \Varid{n}\mathbin{>}\mathrm{0}{}\<[26]%
\>[26]{}\mathrel{=}{}\<[26E]%
\>[29]{}\Varid{aux}\;(\Varid{knuth}_{1\mathrm{b}}\;\Varid{a}\;(\Varid{n}\mathbin{-}\mathrm{1})){}\<[E]%
\\
\>[3]{}\hsindent{2}{}\<[5]%
\>[5]{}\mathbf{where}\;{}\<[12]%
\>[12]{}\Varid{aux}\;{}\<[17]%
\>[17]{}\Varid{f}\;{}\<[20]%
\>[20]{}\mathrm{0}{}\<[31]%
\>[31]{}\mathrel{=}{}\<[31E]%
\>[34]{}\mathrm{1}{}\<[E]%
\\
\>[12]{}\Varid{aux}\;{}\<[17]%
\>[17]{}\Varid{f}\;{}\<[20]%
\>[20]{}\Varid{b}\mid \Varid{b}\mathbin{>}\mathrm{0}{}\<[31]%
\>[31]{}\mathrel{=}{}\<[31E]%
\>[34]{}\Varid{f}\;(\Varid{aux}\;\Varid{f}\;(\Varid{b}\mathbin{-}\mathrm{1})){}\<[E]%
\ColumnHook
\end{hscode}\resethooks

Now we may invoke the universal property of the fold operator to
abstract from the outer recursion first, as in the previous example.
\begin{hscode}\SaveRestoreHook
\column{B}{@{}>{\hspre}l<{\hspost}@{}}%
\column{3}{@{}>{\hspre}l<{\hspost}@{}}%
\column{5}{@{}>{\hspre}l<{\hspost}@{}}%
\column{12}{@{}>{\hspre}l<{\hspost}@{}}%
\column{17}{@{}>{\hspre}l<{\hspost}@{}}%
\column{20}{@{}>{\hspre}l<{\hspost}@{}}%
\column{31}{@{}>{\hspre}c<{\hspost}@{}}%
\column{31E}{@{}l@{}}%
\column{34}{@{}>{\hspre}l<{\hspost}@{}}%
\column{E}{@{}>{\hspre}l<{\hspost}@{}}%
\>[3]{}\Varid{knuth}_{2\mathrm{a}}\;\Varid{a}\mathrel{=}\Varid{foldn}\;\Varid{aux}\;(\Varid{a}\mathbin{*}){}\<[E]%
\\
\>[3]{}\hsindent{2}{}\<[5]%
\>[5]{}\mathbf{where}\;{}\<[12]%
\>[12]{}\Varid{aux}\;{}\<[17]%
\>[17]{}\Varid{f}\;{}\<[20]%
\>[20]{}\mathrm{0}{}\<[31]%
\>[31]{}\mathrel{=}{}\<[31E]%
\>[34]{}\mathrm{1}{}\<[E]%
\\
\>[12]{}\Varid{aux}\;{}\<[17]%
\>[17]{}\Varid{f}\;{}\<[20]%
\>[20]{}\Varid{b}\mid \Varid{b}\mathbin{>}\mathrm{0}{}\<[31]%
\>[31]{}\mathrel{=}{}\<[31E]%
\>[34]{}\Varid{f}\;(\Varid{aux}\;\Varid{f}\;(\Varid{b}\mathbin{-}\mathrm{1})){}\<[E]%
\ColumnHook
\end{hscode}\resethooks
Or, alternatively, we may abstract from the inner recursion first.
\begin{hscode}\SaveRestoreHook
\column{B}{@{}>{\hspre}l<{\hspost}@{}}%
\column{3}{@{}>{\hspre}l<{\hspost}@{}}%
\column{24}{@{}>{\hspre}c<{\hspost}@{}}%
\column{24E}{@{}l@{}}%
\column{27}{@{}>{\hspre}l<{\hspost}@{}}%
\column{E}{@{}>{\hspre}l<{\hspost}@{}}%
\>[3]{}\Varid{knuth}_{2\mathrm{b}}\;\Varid{a}\;\mathrm{0}{}\<[24]%
\>[24]{}\mathrel{=}{}\<[24E]%
\>[27]{}(\Varid{a}\mathbin{*}){}\<[E]%
\\
\>[3]{}\Varid{knuth}_{2\mathrm{b}}\;\Varid{a}\;\Varid{n}\mid \Varid{n}\mathbin{>}\mathrm{0}{}\<[24]%
\>[24]{}\mathrel{=}{}\<[24E]%
\>[27]{}(\lambda \Varid{f}\to \Varid{foldn}\;\Varid{f}\;\mathrm{1})\;(\Varid{knuth}_{2\mathrm{b}}\;\Varid{a}\;(\Varid{n}\mathbin{-}\mathrm{1})){}\<[E]%
\ColumnHook
\end{hscode}\resethooks
Either way, double recursion abstraction leads to:
\begin{hscode}\SaveRestoreHook
\column{B}{@{}>{\hspre}l<{\hspost}@{}}%
\column{3}{@{}>{\hspre}l<{\hspost}@{}}%
\column{E}{@{}>{\hspre}l<{\hspost}@{}}%
\>[3]{}\Varid{knuth}_3\;\Varid{a}\mathrel{=}\Varid{foldn}\;(\lambda \Varid{f}\to \Varid{foldn}\;\Varid{f}\;\mathrm{1})\;(\Varid{a}\mathbin{*}){}\<[E]%
\ColumnHook
\end{hscode}\resethooks
For the friends of point-free humor this is:
\begin{hscode}\SaveRestoreHook
\column{B}{@{}>{\hspre}l<{\hspost}@{}}%
\column{3}{@{}>{\hspre}l<{\hspost}@{}}%
\column{E}{@{}>{\hspre}l<{\hspost}@{}}%
\>[3]{}\Varid{knuth}_4\mathrel{=}\Varid{foldn}\;(\lambda \Varid{f}\to \Varid{foldn}\;\Varid{f}\;\mathrm{1})\mathbin{\circ}(\mathbin{*}){}\<[E]%
\ColumnHook
\end{hscode}\resethooks

Visual type inference tells us that the nesting pattern is the same as
for the Ackermann function, a first-order instance of fold nested
within a second-order instance.
\begin{hscode}\SaveRestoreHook
\column{B}{@{}>{\hspre}l<{\hspost}@{}}%
\column{3}{@{}>{\hspre}l<{\hspost}@{}}%
\column{E}{@{}>{\hspre}l<{\hspost}@{}}%
\>[3]{}\Varid{knuth}_5\mathrel{=}\Varid{foldn}_2\;(\lambda \Varid{f}\to \Varid{foldn}_1\;\Varid{f}\;\mathrm{1})\mathbin{\circ}(\mathbin{*}){}\<[E]%
\ColumnHook
\end{hscode}\resethooks
\subsection{Verification}

Although the mode of derivation has paralleled the, already verified,
Ackermann function quite closely, it might be assuring to repeat the
verification process for Knuth's arrows.  As the base case for \ensuremath{\Varid{n}} we
have
\begin{hscode}\SaveRestoreHook
\column{B}{@{}>{\hspre}l<{\hspost}@{}}%
\column{3}{@{}>{\hspre}l<{\hspost}@{}}%
\column{17}{@{}>{\hspre}c<{\hspost}@{}}%
\column{17E}{@{}l@{}}%
\column{21}{@{}>{\hspre}l<{\hspost}@{}}%
\column{E}{@{}>{\hspre}l<{\hspost}@{}}%
\>[3]{}\Varid{knuth}_3\;\Varid{a}\;\mathrm{0}\;\Varid{b}{}\<[17]%
\>[17]{}\equiv {}\<[17E]%
\>[21]{}\Varid{foldn}\;(\lambda \Varid{f}\to \Varid{foldn}\;\Varid{f}\;\mathrm{1})\;(\Varid{a}\mathbin{*})\;\mathrm{0}\;\Varid{b}{}\<[E]%
\\
\>[17]{}\equiv {}\<[17E]%
\>[21]{}(\Varid{a}\mathbin{*})\;\Varid{b}{}\<[E]%
\\
\>[17]{}\equiv {}\<[17E]%
\>[21]{}\Varid{a}\mathbin{*}\Varid{b}{}\<[E]%
\ColumnHook
\end{hscode}\resethooks
and as the inductive case we have:
\begin{hscode}\SaveRestoreHook
\column{B}{@{}>{\hspre}l<{\hspost}@{}}%
\column{3}{@{}>{\hspre}l<{\hspost}@{}}%
\column{23}{@{}>{\hspre}c<{\hspost}@{}}%
\column{23E}{@{}l@{}}%
\column{27}{@{}>{\hspre}l<{\hspost}@{}}%
\column{49}{@{}>{\hspre}l<{\hspost}@{}}%
\column{81}{@{}>{\hspre}c<{\hspost}@{}}%
\column{81E}{@{}l@{}}%
\column{84}{@{}>{\hspre}l<{\hspost}@{}}%
\column{E}{@{}>{\hspre}l<{\hspost}@{}}%
\>[3]{}\Varid{knuth}_3\;\Varid{a}\;\Varid{n}\mid \Varid{n}\mathbin{>}\mathrm{0}{}\<[23]%
\>[23]{}\equiv {}\<[23E]%
\>[49]{}\Varid{foldn}\;(\lambda \Varid{f}\to \Varid{foldn}\;\Varid{f}\;\mathrm{1})\;(\Varid{a}\mathbin{*})\;{}\<[84]%
\>[84]{}\Varid{n}{}\<[E]%
\\
\>[23]{}\equiv {}\<[23E]%
\>[27]{}(\lambda \Varid{f}\to \Varid{foldn}\;\Varid{f}\;\mathrm{1})\;({}\<[49]%
\>[49]{}\Varid{foldn}\;(\lambda \Varid{f}\to \Varid{foldn}\;\Varid{f}\;\mathrm{1})\;(\Varid{a}\mathbin{*})\;{}\<[81]%
\>[81]{}({}\<[81E]%
\>[84]{}\Varid{n}\mathbin{-}\mathrm{1})){}\<[E]%
\\
\>[23]{}\equiv {}\<[23E]%
\>[27]{}(\lambda \Varid{f}\to \Varid{foldn}\;\Varid{f}\;\mathrm{1})\;({}\<[49]%
\>[49]{}\Varid{knuth}_3\;\Varid{a}\;{}\<[81]%
\>[81]{}({}\<[81E]%
\>[84]{}\Varid{n}\mathbin{-}\mathrm{1})){}\<[E]%
\\
\>[23]{}\equiv {}\<[23E]%
\>[27]{}\Varid{foldn}\;(\Varid{knuth}_3\;\Varid{a}\;(\Varid{n}\mathbin{-}\mathrm{1}))\;\mathrm{1}{}\<[E]%
\ColumnHook
\end{hscode}\resethooks
As the base case for \ensuremath{\Varid{b}} we have
\begin{hscode}\SaveRestoreHook
\column{B}{@{}>{\hspre}l<{\hspost}@{}}%
\column{3}{@{}>{\hspre}l<{\hspost}@{}}%
\column{17}{@{}>{\hspre}c<{\hspost}@{}}%
\column{17E}{@{}l@{}}%
\column{21}{@{}>{\hspre}l<{\hspost}@{}}%
\column{E}{@{}>{\hspre}l<{\hspost}@{}}%
\>[3]{}\Varid{knuth}_3\;\Varid{a}\;\Varid{n}\;\mathrm{0}{}\<[17]%
\>[17]{}\equiv {}\<[17E]%
\>[21]{}\Varid{foldn}\;(\Varid{knuth}_3\;\Varid{a}\;(\Varid{n}\mathbin{-}\mathrm{1}))\;\mathrm{1}\;\mathrm{0}{}\<[E]%
\\
\>[17]{}\equiv {}\<[17E]%
\>[21]{}\mathrm{1}{}\<[E]%
\ColumnHook
\end{hscode}\resethooks
and as the inductive case we have:
\begin{hscode}\SaveRestoreHook
\column{B}{@{}>{\hspre}l<{\hspost}@{}}%
\column{3}{@{}>{\hspre}l<{\hspost}@{}}%
\column{25}{@{}>{\hspre}c<{\hspost}@{}}%
\column{25E}{@{}l@{}}%
\column{29}{@{}>{\hspre}l<{\hspost}@{}}%
\column{49}{@{}>{\hspre}l<{\hspost}@{}}%
\column{77}{@{}>{\hspre}c<{\hspost}@{}}%
\column{77E}{@{}l@{}}%
\column{80}{@{}>{\hspre}l<{\hspost}@{}}%
\column{E}{@{}>{\hspre}l<{\hspost}@{}}%
\>[3]{}\Varid{knuth}_3\;\Varid{a}\;\Varid{n}\;\Varid{b}\mid \Varid{b}\mathbin{>}\mathrm{0}{}\<[25]%
\>[25]{}\equiv {}\<[25E]%
\>[49]{}\Varid{foldn}\;(\Varid{knuth}_3\;\Varid{a}\;(\Varid{n}\mathbin{-}\mathrm{1}))\;\mathrm{1}\;{}\<[80]%
\>[80]{}\Varid{b}{}\<[E]%
\\
\>[25]{}\equiv {}\<[25E]%
\>[29]{}\Varid{knuth}_3\;\Varid{a}\;(\Varid{n}\mathbin{-}\mathrm{1})\;({}\<[49]%
\>[49]{}\Varid{foldn}\;(\Varid{knuth}_3\;\Varid{a}\;(\Varid{n}\mathbin{-}\mathrm{1}))\;\mathrm{1}\;{}\<[77]%
\>[77]{}({}\<[77E]%
\>[80]{}\Varid{b}\mathbin{-}\mathrm{1})){}\<[E]%
\\
\>[25]{}\equiv {}\<[25E]%
\>[29]{}\Varid{knuth}_3\;\Varid{a}\;(\Varid{n}\mathbin{-}\mathrm{1})\;({}\<[49]%
\>[49]{}\Varid{knuth}_3\;\Varid{a}\;\Varid{n}\;{}\<[77]%
\>[77]{}({}\<[77E]%
\>[80]{}\Varid{b}\mathbin{-}\mathrm{1})){}\<[E]%
\ColumnHook
\end{hscode}\resethooks
$\Box$

\pagebreak
\section{Conway's Chained Arrows}

Knuth's up-arrow notation was conceived as an iterative generalization of
the step from addition to multiplication and from there to
exponentiation, remaining in the realm of binary operators.  Conway's
chained arrow notation releases the latter restriction to encode even
huger numbers.  It defines a single operator on a list of arbitrarily
many numbers, traditionally called ``chains'' and written with interspersed arrows (hence
the name).  A typical self-referential definition is the following
system of equations, where $p, q$ are numbers and $X$ is a nonempty
subsequence:
\begin{equation*}
  \begin{array}{rcl@{\qquad}r}
  \langle p\rangle &=& p
  \\
  \langle p \to q\rangle &=& p^q
  \\
  \langle X \to p \to 1\rangle &=& \langle X \to p\rangle
  \\
  \langle X \to 1 \to q\rangle &=& \langle X \to 1\rangle
  \\
  \langle X \to p \to q\rangle &=& \bigl\langle X \to \langle X \to (p - 1) \to q\rangle \to (q - 1)\bigr\rangle
           & (p, q > 1)
  \\ 
  \multicolumn{4}{l}{\qquad\textrm{with the canonical extension}}
  \\
  \langle\,\rangle &=& 1
  \end{array}
\end{equation*}

We have added angled brackets around chains to emphasize the relevance
of bracketing: since a chain denotes a number, chains may be nested
(as in the last equation); but unlike for many other arithmetic
operators, a chain of more than two elements is not \emph{merely} an
abbreviation for the nested iteration of a binary operator $\to$.  The
full situation is complicated: even though generally $\langle o \to p
\to q\rangle$ is different from both $\langle\langle o \to p \rangle
\to q \rangle$ and $\langle o \to \langle p \to q \rangle \rangle$,
this does not imply that the overall operation \emph{cannot} be
expressed by a fold operator acting binarily on the list of
numbers---that will be exactly our next move.

\subsection{Derivation}

The above equations, read left-to-right, specify a total evaluation
function.  It can be translated to a Haskell function on lists of
strictly positive numbers
\begin{hscode}\SaveRestoreHook
\column{B}{@{}>{\hspre}l<{\hspost}@{}}%
\column{3}{@{}>{\hspre}l<{\hspost}@{}}%
\column{5}{@{}>{\hspre}l<{\hspost}@{}}%
\column{15}{@{}>{\hspre}c<{\hspost}@{}}%
\column{15E}{@{}l@{}}%
\column{18}{@{}>{\hspre}l<{\hspost}@{}}%
\column{21}{@{}>{\hspre}c<{\hspost}@{}}%
\column{21E}{@{}l@{}}%
\column{24}{@{}>{\hspre}l<{\hspost}@{}}%
\column{46}{@{}>{\hspre}c<{\hspost}@{}}%
\column{46E}{@{}l@{}}%
\column{49}{@{}>{\hspre}l<{\hspost}@{}}%
\column{67}{@{}>{\hspre}c<{\hspost}@{}}%
\column{67E}{@{}l@{}}%
\column{70}{@{}>{\hspre}l<{\hspost}@{}}%
\column{79}{@{}>{\hspre}c<{\hspost}@{}}%
\column{79E}{@{}l@{}}%
\column{82}{@{}>{\hspre}l<{\hspost}@{}}%
\column{E}{@{}>{\hspre}l<{\hspost}@{}}%
\>[3]{}\Varid{conway}_0\;[\mskip1.5mu \mskip1.5mu]{}\<[46]%
\>[46]{}\mathrel{=}{}\<[46E]%
\>[49]{}\mathrm{1}{}\<[E]%
\\
\>[3]{}\Varid{conway}_0\;[\mskip1.5mu \Varid{p}\mskip1.5mu]{}\<[46]%
\>[46]{}\mathrel{=}{}\<[46E]%
\>[49]{}\Varid{p}{}\<[E]%
\\
\>[3]{}\Varid{conway}_0\;[\mskip1.5mu \Varid{q},\Varid{p}\mskip1.5mu]{}\<[46]%
\>[46]{}\mathrel{=}{}\<[46E]%
\>[49]{}\Varid{p}\mathbin{\hat{\enspace}}\Varid{q}{}\<[E]%
\\
\>[3]{}\Varid{conway}_0\;(\mathrm{1}{}\<[15]%
\>[15]{}\mathbin{:}{}\<[15E]%
\>[18]{}\Varid{p}{}\<[21]%
\>[21]{}\mathbin{:}{}\<[21E]%
\>[24]{}\Varid{xs}){}\<[46]%
\>[46]{}\mathrel{=}{}\<[46E]%
\>[49]{}\Varid{conway}_0\;(\Varid{p}\mathbin{:}\Varid{xs}){}\<[E]%
\\
\>[3]{}\Varid{conway}_0\;(\Varid{q}{}\<[15]%
\>[15]{}\mathbin{:}{}\<[15E]%
\>[18]{}\mathrm{1}{}\<[21]%
\>[21]{}\mathbin{:}{}\<[21E]%
\>[24]{}\Varid{xs}){}\<[46]%
\>[46]{}\mathrel{=}{}\<[46E]%
\>[49]{}\Varid{conway}_0\;(\mathrm{1}\mathbin{:}\Varid{xs}){}\<[E]%
\\
\>[3]{}\Varid{conway}_0\;(\Varid{q}{}\<[15]%
\>[15]{}\mathbin{:}{}\<[15E]%
\>[18]{}\Varid{p}{}\<[21]%
\>[21]{}\mathbin{:}{}\<[21E]%
\>[24]{}\Varid{xs})\mid \Varid{p}\mathbin{>}\mathrm{1}\mathrel{\wedge}\Varid{q}\mathbin{>}\mathrm{1}{}\<[46]%
\>[46]{}\mathrel{=}{}\<[46E]%
\>[49]{}\Varid{conway}_0\;((\Varid{q}\mathbin{-}\mathrm{1}){}\<[67]%
\>[67]{}\mathbin{:}{}\<[67E]%
\>[70]{}\Varid{p'}{}\<[79]%
\>[79]{}\mathbin{:}{}\<[79E]%
\>[82]{}\Varid{xs}){}\<[E]%
\\
\>[3]{}\hsindent{2}{}\<[5]%
\>[5]{}\mathbf{where}\;\Varid{p'}{}\<[46]%
\>[46]{}\mathrel{=}{}\<[46E]%
\>[49]{}\Varid{conway}_0\;(\Varid{q}{}\<[67]%
\>[67]{}\mathbin{:}{}\<[67E]%
\>[70]{}(\Varid{p}\mathbin{-}\mathrm{1}){}\<[79]%
\>[79]{}\mathbin{:}{}\<[79E]%
\>[82]{}\Varid{xs}){}\<[E]%
\ColumnHook
\end{hscode}\resethooks
where the order is reversed, because Conway's notation matches from
the right, but Haskell lists decompose from the left.  In order to
apply recursion abstraction successfully to this messy function, its
arguments need to be regularized and reordered in a particular way.
Firstly, we note that the first two cases, for lists of length less
than two, are special, in the sense that they are never reached by
recursion; the third case, for lists of length two exactly, is the
actual recursive base case.  

Secondly, we notice that the remaining cases all involve two or more
numbers, and the nested recursion occurs quite inconveniently in
second position in the list.  Hence we separate a non-recursive
front-end function for the first two cases
\begin{hscode}\SaveRestoreHook
\column{B}{@{}>{\hspre}l<{\hspost}@{}}%
\column{3}{@{}>{\hspre}l<{\hspost}@{}}%
\column{12}{@{}>{\hspre}l<{\hspost}@{}}%
\column{15}{@{}>{\hspre}l<{\hspost}@{}}%
\column{29}{@{}>{\hspre}c<{\hspost}@{}}%
\column{29E}{@{}l@{}}%
\column{32}{@{}>{\hspre}l<{\hspost}@{}}%
\column{E}{@{}>{\hspre}l<{\hspost}@{}}%
\>[3]{}\Varid{cfront}_1\;{}\<[12]%
\>[12]{}\anonymous \;{}\<[15]%
\>[15]{}[\mskip1.5mu \mskip1.5mu]{}\<[29]%
\>[29]{}\mathrel{=}{}\<[29E]%
\>[32]{}\mathrm{1}{}\<[E]%
\\
\>[3]{}\Varid{cfront}_1\;{}\<[12]%
\>[12]{}\anonymous \;{}\<[15]%
\>[15]{}[\mskip1.5mu \Varid{p}\mskip1.5mu]{}\<[29]%
\>[29]{}\mathrel{=}{}\<[29E]%
\>[32]{}\Varid{p}{}\<[E]%
\\
\>[3]{}\Varid{cfront}_1\;{}\<[12]%
\>[12]{}\Varid{b}\;{}\<[15]%
\>[15]{}(\Varid{q}\mathbin{:}\Varid{p}\mathbin{:}\Varid{xs}){}\<[29]%
\>[29]{}\mathrel{=}{}\<[29E]%
\>[32]{}\Varid{b}\;\Varid{xs}\;\Varid{q}\;\Varid{p}{}\<[E]%
\ColumnHook
\end{hscode}\resethooks
and a recursive back-end function, with the first two list elements
expanded and reordered, for the other cases
\begin{hscode}\SaveRestoreHook
\column{B}{@{}>{\hspre}l<{\hspost}@{}}%
\column{3}{@{}>{\hspre}l<{\hspost}@{}}%
\column{6}{@{}>{\hspre}l<{\hspost}@{}}%
\column{11}{@{}>{\hspre}l<{\hspost}@{}}%
\column{21}{@{}>{\hspre}l<{\hspost}@{}}%
\column{24}{@{}>{\hspre}l<{\hspost}@{}}%
\column{44}{@{}>{\hspre}c<{\hspost}@{}}%
\column{44E}{@{}l@{}}%
\column{47}{@{}>{\hspre}l<{\hspost}@{}}%
\column{55}{@{}>{\hspre}l<{\hspost}@{}}%
\column{59}{@{}>{\hspre}l<{\hspost}@{}}%
\column{68}{@{}>{\hspre}l<{\hspost}@{}}%
\column{E}{@{}>{\hspre}l<{\hspost}@{}}%
\>[3]{}\Varid{cback}_1\;{}\<[11]%
\>[11]{}[\mskip1.5mu \mskip1.5mu]\;{}\<[21]%
\>[21]{}\Varid{q}\;{}\<[24]%
\>[24]{}\Varid{p}{}\<[44]%
\>[44]{}\mathrel{=}{}\<[44E]%
\>[47]{}\Varid{p}\mathbin{\hat{\enspace}}\Varid{q}{}\<[E]%
\\
\>[3]{}\Varid{cback}_1\;{}\<[11]%
\>[11]{}(\Varid{o}\mathbin{:}\Varid{ys})\;{}\<[21]%
\>[21]{}\mathrm{1}\;{}\<[24]%
\>[24]{}\Varid{p}{}\<[44]%
\>[44]{}\mathrel{=}{}\<[44E]%
\>[47]{}\Varid{cback}_1\;{}\<[55]%
\>[55]{}\Varid{ys}\;{}\<[59]%
\>[59]{}\Varid{p}\;{}\<[68]%
\>[68]{}\Varid{o}{}\<[E]%
\\
\>[3]{}\Varid{cback}_1\;{}\<[11]%
\>[11]{}(\Varid{o}\mathbin{:}\Varid{ys})\;{}\<[21]%
\>[21]{}\Varid{q}\;{}\<[24]%
\>[24]{}\mathrm{1}{}\<[44]%
\>[44]{}\mathrel{=}{}\<[44E]%
\>[47]{}\Varid{cback}_1\;{}\<[55]%
\>[55]{}\Varid{ys}\;{}\<[59]%
\>[59]{}\mathrm{1}\;{}\<[68]%
\>[68]{}\Varid{o}{}\<[E]%
\\
\>[3]{}\Varid{cback}_1\;{}\<[11]%
\>[11]{}\Varid{xs}\;{}\<[21]%
\>[21]{}\Varid{q}\;{}\<[24]%
\>[24]{}\Varid{p}\mid \Varid{p}\mathbin{>}\mathrm{1}\mathrel{\wedge}\Varid{q}\mathbin{>}\mathrm{1}{}\<[44]%
\>[44]{}\mathrel{=}{}\<[44E]%
\>[47]{}\Varid{cback}_1\;{}\<[55]%
\>[55]{}\Varid{xs}\;{}\<[59]%
\>[59]{}(\Varid{q}\mathbin{-}\mathrm{1})\;{}\<[68]%
\>[68]{}\Varid{p'}{}\<[E]%
\\
\>[3]{}\hsindent{3}{}\<[6]%
\>[6]{}\mathbf{where}\;\Varid{p'}{}\<[44]%
\>[44]{}\mathrel{=}{}\<[44E]%
\>[47]{}\Varid{cback}_1\;{}\<[55]%
\>[55]{}\Varid{xs}\;{}\<[59]%
\>[59]{}\Varid{q}\;{}\<[68]%
\>[68]{}(\Varid{p}\mathbin{-}\mathrm{1}){}\<[E]%
\ColumnHook
\end{hscode}\resethooks
such that
\begin{hscode}\SaveRestoreHook
\column{B}{@{}>{\hspre}l<{\hspost}@{}}%
\column{3}{@{}>{\hspre}l<{\hspost}@{}}%
\column{E}{@{}>{\hspre}l<{\hspost}@{}}%
\>[3]{}\Varid{conway}\equiv \Varid{cfront}\;\Varid{cback}{}\<[E]%
\ColumnHook
\end{hscode}\resethooks
Note that the reordering of arguments, however unintuitive at first
glance, moves the nested recursion \ensuremath{\Varid{p'}} from the awkward middle to a
more manageable final position.

Conway's arrows are traditionally defined starting from one rather
than zero.  But unlike for Knuth's arrows, there is no simple
extension to zero arguments.  In order to make the recursion pattern
more similar to the previous ones, we consider a variant where all
argument numbers are reduced by one in the front-end.\footnote{The notation 
\ensuremath{(-\;\mathrm{1})} is actually \ensuremath{(\Varid{subtract}\;\mathrm{1})} in Haskell.}
\begin{hscode}\SaveRestoreHook
\column{B}{@{}>{\hspre}l<{\hspost}@{}}%
\column{3}{@{}>{\hspre}l<{\hspost}@{}}%
\column{12}{@{}>{\hspre}l<{\hspost}@{}}%
\column{15}{@{}>{\hspre}l<{\hspost}@{}}%
\column{21}{@{}>{\hspre}l<{\hspost}@{}}%
\column{38}{@{}>{\hspre}c<{\hspost}@{}}%
\column{38E}{@{}l@{}}%
\column{41}{@{}>{\hspre}l<{\hspost}@{}}%
\column{E}{@{}>{\hspre}l<{\hspost}@{}}%
\>[3]{}\Varid{cfront}_2\;{}\<[12]%
\>[12]{}\anonymous \;{}\<[15]%
\>[15]{}[\mskip1.5mu \mskip1.5mu]{}\<[38]%
\>[38]{}\mathrel{=}{}\<[38E]%
\>[41]{}\mathrm{1}{}\<[E]%
\\
\>[3]{}\Varid{cfront}_2\;{}\<[12]%
\>[12]{}\anonymous \;{}\<[15]%
\>[15]{}[\mskip1.5mu \Varid{p}\mskip1.5mu]{}\<[38]%
\>[38]{}\mathrel{=}{}\<[38E]%
\>[41]{}\Varid{p}{}\<[E]%
\\
\>[3]{}\Varid{cfront}_2\;{}\<[12]%
\>[12]{}\Varid{b}\;{}\<[15]%
\>[15]{}\Varid{xs}{}\<[21]%
\>[21]{}\mid \Varid{length}\;\Varid{xs}\mathbin{>}\mathrm{1}{}\<[38]%
\>[38]{}\mathrel{=}{}\<[38E]%
\>[41]{}\Varid{cfront}_1\;\Varid{b}\;(\Varid{map}\;(-\;\mathrm{1})\;\Varid{xs}){}\<[E]%
\ColumnHook
\end{hscode}\resethooks
This gives us a slightly modified back-end:
\begin{hscode}\SaveRestoreHook
\column{B}{@{}>{\hspre}l<{\hspost}@{}}%
\column{3}{@{}>{\hspre}l<{\hspost}@{}}%
\column{6}{@{}>{\hspre}l<{\hspost}@{}}%
\column{11}{@{}>{\hspre}l<{\hspost}@{}}%
\column{21}{@{}>{\hspre}l<{\hspost}@{}}%
\column{24}{@{}>{\hspre}l<{\hspost}@{}}%
\column{44}{@{}>{\hspre}c<{\hspost}@{}}%
\column{44E}{@{}l@{}}%
\column{47}{@{}>{\hspre}l<{\hspost}@{}}%
\column{55}{@{}>{\hspre}l<{\hspost}@{}}%
\column{59}{@{}>{\hspre}l<{\hspost}@{}}%
\column{68}{@{}>{\hspre}l<{\hspost}@{}}%
\column{73}{@{}>{\hspre}c<{\hspost}@{}}%
\column{73E}{@{}l@{}}%
\column{76}{@{}>{\hspre}l<{\hspost}@{}}%
\column{E}{@{}>{\hspre}l<{\hspost}@{}}%
\>[3]{}\Varid{cback}_2\;{}\<[11]%
\>[11]{}[\mskip1.5mu \mskip1.5mu]\;{}\<[21]%
\>[21]{}\Varid{q}\;{}\<[24]%
\>[24]{}\Varid{p}{}\<[44]%
\>[44]{}\mathrel{=}{}\<[44E]%
\>[47]{}(\Varid{p}\mathbin{+}\mathrm{1})\mathbin{\hat{\enspace}}(\Varid{q}\mathbin{+}\mathrm{1}){}\<[E]%
\\
\>[3]{}\Varid{cback}_2\;{}\<[11]%
\>[11]{}(\Varid{o}\mathbin{:}\Varid{ys})\;{}\<[21]%
\>[21]{}\mathrm{0}\;{}\<[24]%
\>[24]{}\Varid{p}{}\<[44]%
\>[44]{}\mathrel{=}{}\<[44E]%
\>[47]{}\Varid{cback}_2\;{}\<[55]%
\>[55]{}\Varid{ys}\;{}\<[59]%
\>[59]{}\Varid{p}\;{}\<[68]%
\>[68]{}\Varid{o}{}\<[E]%
\\
\>[3]{}\Varid{cback}_2\;{}\<[11]%
\>[11]{}(\Varid{o}\mathbin{:}\Varid{ys})\;{}\<[21]%
\>[21]{}\Varid{q}\;{}\<[24]%
\>[24]{}\mathrm{0}{}\<[44]%
\>[44]{}\mathrel{=}{}\<[44E]%
\>[47]{}\Varid{cback}_2\;{}\<[55]%
\>[55]{}\Varid{ys}\;{}\<[59]%
\>[59]{}\mathrm{0}\;{}\<[68]%
\>[68]{}\Varid{o}{}\<[E]%
\\
\>[3]{}\Varid{cback}_2\;{}\<[11]%
\>[11]{}\Varid{xs}\;{}\<[21]%
\>[21]{}\Varid{q}\;{}\<[24]%
\>[24]{}\Varid{p}\mid \Varid{p}\mathbin{>}\mathrm{0}\mathrel{\wedge}\Varid{q}\mathbin{>}\mathrm{0}{}\<[44]%
\>[44]{}\mathrel{=}{}\<[44E]%
\>[47]{}\Varid{cback}_2\;{}\<[55]%
\>[55]{}\Varid{xs}\;{}\<[59]%
\>[59]{}(\Varid{q}\mathbin{-}\mathrm{1})\;{}\<[68]%
\>[68]{}(\Varid{p'}{}\<[73]%
\>[73]{}\mathbin{-}{}\<[73E]%
\>[76]{}\mathrm{1}){}\<[E]%
\\
\>[3]{}\hsindent{3}{}\<[6]%
\>[6]{}\mathbf{where}\;\Varid{p'}{}\<[44]%
\>[44]{}\mathrel{=}{}\<[44E]%
\>[47]{}\Varid{cback}_2\;{}\<[55]%
\>[55]{}\Varid{xs}\;{}\<[59]%
\>[59]{}\Varid{q}\;{}\<[68]%
\>[68]{}(\Varid{p}{}\<[73]%
\>[73]{}\mathbin{-}{}\<[73E]%
\>[76]{}\mathrm{1}){}\<[E]%
\ColumnHook
\end{hscode}\resethooks

Note the two corrections \ensuremath{(\mathbin{+}\mathrm{1})} where arguments flow to results, and the single correction \ensuremath{(-\;\mathrm{1})} where a recursive result flows back to an argument.

The first argument is the technical novelty of this exercise, because it is of list type.  We tackle it by double currying:
           \begin{hscode}\SaveRestoreHook
\column{B}{@{}>{\hspre}l<{\hspost}@{}}%
\column{3}{@{}>{\hspre}l<{\hspost}@{}}%
\column{5}{@{}>{\hspre}l<{\hspost}@{}}%
\column{11}{@{}>{\hspre}l<{\hspost}@{}}%
\column{12}{@{}>{\hspre}l<{\hspost}@{}}%
\column{14}{@{}>{\hspre}l<{\hspost}@{}}%
\column{17}{@{}>{\hspre}l<{\hspost}@{}}%
\column{20}{@{}>{\hspre}l<{\hspost}@{}}%
\column{24}{@{}>{\hspre}c<{\hspost}@{}}%
\column{24E}{@{}l@{}}%
\column{27}{@{}>{\hspre}l<{\hspost}@{}}%
\column{40}{@{}>{\hspre}c<{\hspost}@{}}%
\column{40E}{@{}l@{}}%
\column{43}{@{}>{\hspre}l<{\hspost}@{}}%
\column{51}{@{}>{\hspre}l<{\hspost}@{}}%
\column{55}{@{}>{\hspre}l<{\hspost}@{}}%
\column{64}{@{}>{\hspre}l<{\hspost}@{}}%
\column{69}{@{}>{\hspre}c<{\hspost}@{}}%
\column{69E}{@{}l@{}}%
\column{72}{@{}>{\hspre}l<{\hspost}@{}}%
\column{E}{@{}>{\hspre}l<{\hspost}@{}}%
\>[3]{}\Varid{cback}_3\;{}\<[11]%
\>[11]{}[\mskip1.5mu \mskip1.5mu]{}\<[24]%
\>[24]{}\mathrel{=}{}\<[24E]%
\>[27]{}\Varid{cpow}{}\<[E]%
\\
\>[3]{}\Varid{cback}_3\;{}\<[11]%
\>[11]{}\Varid{xs}\mathord{@}(\Varid{o}\mathbin{:}\Varid{ys}){}\<[24]%
\>[24]{}\mathrel{=}{}\<[24E]%
\>[27]{}\Varid{aux}{}\<[E]%
\\
\>[3]{}\hsindent{2}{}\<[5]%
\>[5]{}\mathbf{where}\;{}\<[12]%
\>[12]{}\Varid{aux}\;{}\<[17]%
\>[17]{}\mathrm{0}\;{}\<[20]%
\>[20]{}\Varid{p}{}\<[40]%
\>[40]{}\mathrel{=}{}\<[40E]%
\>[43]{}\Varid{cback}_3\;{}\<[51]%
\>[51]{}\Varid{ys}\;{}\<[55]%
\>[55]{}\Varid{p}\;{}\<[64]%
\>[64]{}\Varid{o}{}\<[E]%
\\
\>[12]{}\Varid{aux}\;{}\<[17]%
\>[17]{}\Varid{q}\;{}\<[20]%
\>[20]{}\mathrm{0}{}\<[40]%
\>[40]{}\mathrel{=}{}\<[40E]%
\>[43]{}\Varid{cback}_3\;{}\<[51]%
\>[51]{}\Varid{ys}\;{}\<[55]%
\>[55]{}\mathrm{0}\;{}\<[64]%
\>[64]{}\Varid{o}{}\<[E]%
\\
\>[12]{}\Varid{aux}\;{}\<[17]%
\>[17]{}\Varid{q}\;{}\<[20]%
\>[20]{}\Varid{p}\mid \Varid{p}\mathbin{>}\mathrm{0}\mathrel{\wedge}\Varid{q}\mathbin{>}\mathrm{0}{}\<[40]%
\>[40]{}\mathrel{=}{}\<[40E]%
\>[43]{}\Varid{cback}_3\;{}\<[51]%
\>[51]{}\Varid{xs}\;{}\<[55]%
\>[55]{}(\Varid{q}\mathbin{-}\mathrm{1})\;{}\<[64]%
\>[64]{}(\Varid{p'}{}\<[69]%
\>[69]{}\mathbin{-}{}\<[69E]%
\>[72]{}\mathrm{1}){}\<[E]%
\\
\>[12]{}\hsindent{2}{}\<[14]%
\>[14]{}\mathbf{where}\;\Varid{p'}{}\<[40]%
\>[40]{}\mathrel{=}{}\<[40E]%
\>[43]{}\Varid{cback}_3\;{}\<[51]%
\>[51]{}\Varid{xs}\;{}\<[55]%
\>[55]{}\Varid{q}\;{}\<[64]%
\>[64]{}(\Varid{p}{}\<[69]%
\>[69]{}\mathbin{-}{}\<[69E]%
\>[72]{}\mathrm{1}){}\<[E]%
\\[\blanklineskip]%
\>[3]{}\Varid{cpow}\;\Varid{q}\;\Varid{p}\mathrel{=}(\Varid{p}\mathbin{+}\mathrm{1})\mathbin{\hat{\enspace}}(\Varid{q}\mathbin{+}\mathrm{1}){}\<[E]%
\ColumnHook
\end{hscode}\resethooks
Note the auxiliary function \ensuremath{\Varid{cpow}}, and the use of \emph{en passant}
pattern matching to unify the different heads of the three recursive cases.
The fold operator for lists takes generators with two arguments, a
non-recursive one for the head and a recursive one for the tail.  We
shape the auxiliary function accordingly by beta expansion,
      \begin{hscode}\SaveRestoreHook
\column{B}{@{}>{\hspre}l<{\hspost}@{}}%
\column{3}{@{}>{\hspre}l<{\hspost}@{}}%
\column{5}{@{}>{\hspre}l<{\hspost}@{}}%
\column{12}{@{}>{\hspre}l<{\hspost}@{}}%
\column{14}{@{}>{\hspre}l<{\hspost}@{}}%
\column{17}{@{}>{\hspre}l<{\hspost}@{}}%
\column{20}{@{}>{\hspre}l<{\hspost}@{}}%
\column{23}{@{}>{\hspre}l<{\hspost}@{}}%
\column{24}{@{}>{\hspre}c<{\hspost}@{}}%
\column{24E}{@{}l@{}}%
\column{26}{@{}>{\hspre}l<{\hspost}@{}}%
\column{27}{@{}>{\hspre}l<{\hspost}@{}}%
\column{46}{@{}>{\hspre}c<{\hspost}@{}}%
\column{46E}{@{}l@{}}%
\column{49}{@{}>{\hspre}l<{\hspost}@{}}%
\column{57}{@{}>{\hspre}l<{\hspost}@{}}%
\column{61}{@{}>{\hspre}l<{\hspost}@{}}%
\column{70}{@{}>{\hspre}l<{\hspost}@{}}%
\column{75}{@{}>{\hspre}c<{\hspost}@{}}%
\column{75E}{@{}l@{}}%
\column{78}{@{}>{\hspre}l<{\hspost}@{}}%
\column{E}{@{}>{\hspre}l<{\hspost}@{}}%
\>[3]{}\Varid{cback}_{3\mathrm{a}}\;[\mskip1.5mu \mskip1.5mu]{}\<[24]%
\>[24]{}\mathrel{=}{}\<[24E]%
\>[27]{}\Varid{cpow}{}\<[E]%
\\
\>[3]{}\Varid{cback}_{3\mathrm{a}}\;\Varid{xs}\mathord{@}(\Varid{o}\mathbin{:}\Varid{ys}){}\<[24]%
\>[24]{}\mathrel{=}{}\<[24E]%
\>[27]{}\Varid{aux}\;\Varid{o}\;(\Varid{cback}_{3\mathrm{a}}\;\Varid{ys}){}\<[E]%
\\
\>[3]{}\hsindent{2}{}\<[5]%
\>[5]{}\mathbf{where}\;{}\<[12]%
\>[12]{}\Varid{aux}\;{}\<[17]%
\>[17]{}\Varid{o}\;{}\<[20]%
\>[20]{}\Varid{k}\;{}\<[23]%
\>[23]{}\mathrm{0}\;{}\<[26]%
\>[26]{}\Varid{p}{}\<[46]%
\>[46]{}\mathrel{=}{}\<[46E]%
\>[49]{}\Varid{k}\;{}\<[61]%
\>[61]{}\Varid{p}\;{}\<[70]%
\>[70]{}\Varid{o}{}\<[E]%
\\
\>[12]{}\Varid{aux}\;{}\<[17]%
\>[17]{}\Varid{o}\;{}\<[20]%
\>[20]{}\Varid{k}\;{}\<[23]%
\>[23]{}\Varid{q}\;{}\<[26]%
\>[26]{}\mathrm{0}{}\<[46]%
\>[46]{}\mathrel{=}{}\<[46E]%
\>[49]{}\Varid{k}\;{}\<[61]%
\>[61]{}\mathrm{0}\;{}\<[70]%
\>[70]{}\Varid{o}{}\<[E]%
\\
\>[12]{}\Varid{aux}\;{}\<[17]%
\>[17]{}\Varid{o}\;{}\<[20]%
\>[20]{}\Varid{k}\;{}\<[23]%
\>[23]{}\Varid{q}\;{}\<[26]%
\>[26]{}\Varid{p}\mid \Varid{p}\mathbin{>}\mathrm{0}\mathrel{\wedge}\Varid{q}\mathbin{>}\mathrm{0}{}\<[46]%
\>[46]{}\mathrel{=}{}\<[46E]%
\>[49]{}\Varid{cback}_3\;{}\<[57]%
\>[57]{}\Varid{xs}\;{}\<[61]%
\>[61]{}(\Varid{q}\mathbin{-}\mathrm{1})\;{}\<[70]%
\>[70]{}(\Varid{p'}{}\<[75]%
\>[75]{}\mathbin{-}{}\<[75E]%
\>[78]{}\mathrm{1}){}\<[E]%
\\
\>[12]{}\hsindent{2}{}\<[14]%
\>[14]{}\mathbf{where}\;\Varid{p'}{}\<[46]%
\>[46]{}\mathrel{=}{}\<[46E]%
\>[49]{}\Varid{cback}_3\;{}\<[57]%
\>[57]{}\Varid{xs}\;{}\<[61]%
\>[61]{}\Varid{q}\;{}\<[70]%
\>[70]{}(\Varid{p}{}\<[75]%
\>[75]{}\mathbin{-}{}\<[75E]%
\>[78]{}\mathrm{1}){}\<[E]%
\ColumnHook
\end{hscode}\resethooks
make the, already familiar, context-sensitive substitution
                   \begin{hscode}\SaveRestoreHook
\column{B}{@{}>{\hspre}l<{\hspost}@{}}%
\column{3}{@{}>{\hspre}l<{\hspost}@{}}%
\column{E}{@{}>{\hspre}l<{\hspost}@{}}%
\>[3]{}\Varid{cback}\;\Varid{xs}\mathord{@}(\Varid{o}\mathbin{:}\Varid{ys})\equiv \Varid{aux}\;\Varid{o}\;(\Varid{cback}\;\Varid{ys})\equiv \Varid{aux}\;\Varid{o}\;\Varid{k}{}\<[E]%
\ColumnHook
\end{hscode}\resethooks
and eliminate the now unused variable \ensuremath{\Varid{xs}} to obtain:\pagebreak
                   \begin{hscode}\SaveRestoreHook
\column{B}{@{}>{\hspre}l<{\hspost}@{}}%
\column{3}{@{}>{\hspre}l<{\hspost}@{}}%
\column{5}{@{}>{\hspre}l<{\hspost}@{}}%
\column{12}{@{}>{\hspre}l<{\hspost}@{}}%
\column{14}{@{}>{\hspre}l<{\hspost}@{}}%
\column{17}{@{}>{\hspre}l<{\hspost}@{}}%
\column{21}{@{}>{\hspre}l<{\hspost}@{}}%
\column{24}{@{}>{\hspre}l<{\hspost}@{}}%
\column{27}{@{}>{\hspre}l<{\hspost}@{}}%
\column{47}{@{}>{\hspre}c<{\hspost}@{}}%
\column{47E}{@{}l@{}}%
\column{50}{@{}>{\hspre}l<{\hspost}@{}}%
\column{59}{@{}>{\hspre}l<{\hspost}@{}}%
\column{68}{@{}>{\hspre}l<{\hspost}@{}}%
\column{73}{@{}>{\hspre}c<{\hspost}@{}}%
\column{73E}{@{}l@{}}%
\column{76}{@{}>{\hspre}l<{\hspost}@{}}%
\column{E}{@{}>{\hspre}l<{\hspost}@{}}%
\>[3]{}\Varid{cback}_{3\mathrm{b}}\;[\mskip1.5mu \mskip1.5mu]{}\<[21]%
\>[21]{}\mathrel{=}{}\<[24]%
\>[24]{}\Varid{cpow}{}\<[E]%
\\
\>[3]{}\Varid{cback}_{3\mathrm{b}}\;(\Varid{o}\mathbin{:}\Varid{ys}){}\<[21]%
\>[21]{}\mathrel{=}{}\<[24]%
\>[24]{}\Varid{aux}\;\Varid{o}\;(\Varid{cback}_{3\mathrm{b}}\;\Varid{ys}){}\<[E]%
\\
\>[3]{}\hsindent{2}{}\<[5]%
\>[5]{}\mathbf{where}\;{}\<[12]%
\>[12]{}\Varid{aux}\;{}\<[17]%
\>[17]{}\Varid{o}\;{}\<[21]%
\>[21]{}\Varid{k}\;{}\<[24]%
\>[24]{}\mathrm{0}\;{}\<[27]%
\>[27]{}\Varid{p}{}\<[47]%
\>[47]{}\mathrel{=}{}\<[47E]%
\>[50]{}\Varid{k}\;{}\<[59]%
\>[59]{}\Varid{p}\;{}\<[68]%
\>[68]{}\Varid{o}{}\<[E]%
\\
\>[12]{}\Varid{aux}\;{}\<[17]%
\>[17]{}\Varid{o}\;{}\<[21]%
\>[21]{}\Varid{k}\;{}\<[24]%
\>[24]{}\Varid{q}\;{}\<[27]%
\>[27]{}\mathrm{0}{}\<[47]%
\>[47]{}\mathrel{=}{}\<[47E]%
\>[50]{}\Varid{k}\;{}\<[59]%
\>[59]{}\mathrm{0}\;{}\<[68]%
\>[68]{}\Varid{o}{}\<[E]%
\\
\>[12]{}\Varid{aux}\;{}\<[17]%
\>[17]{}\Varid{o}\;{}\<[21]%
\>[21]{}\Varid{k}\;{}\<[24]%
\>[24]{}\Varid{q}\;{}\<[27]%
\>[27]{}\Varid{p}\mid \Varid{p}\mathbin{>}\mathrm{0}\mathrel{\wedge}\Varid{q}\mathbin{>}\mathrm{0}{}\<[47]%
\>[47]{}\mathrel{=}{}\<[47E]%
\>[50]{}\Varid{aux}\;\Varid{o}\;\Varid{k}\;{}\<[59]%
\>[59]{}(\Varid{q}\mathbin{-}\mathrm{1})\;{}\<[68]%
\>[68]{}(\Varid{p'}{}\<[73]%
\>[73]{}\mathbin{-}{}\<[73E]%
\>[76]{}\mathrm{1}){}\<[E]%
\\
\>[12]{}\hsindent{2}{}\<[14]%
\>[14]{}\mathbf{where}\;\Varid{p'}{}\<[47]%
\>[47]{}\mathrel{=}{}\<[47E]%
\>[50]{}\Varid{aux}\;\Varid{o}\;\Varid{k}\;{}\<[59]%
\>[59]{}\Varid{q}\;{}\<[68]%
\>[68]{}(\Varid{p}{}\<[73]%
\>[73]{}\mathbin{-}{}\<[73E]%
\>[76]{}\mathrm{1}){}\<[E]%
\ColumnHook
\end{hscode}\resethooks
We readily read off the universal property of list folding and deduce:         
                   \begin{hscode}\SaveRestoreHook
\column{B}{@{}>{\hspre}l<{\hspost}@{}}%
\column{3}{@{}>{\hspre}l<{\hspost}@{}}%
\column{5}{@{}>{\hspre}l<{\hspost}@{}}%
\column{12}{@{}>{\hspre}l<{\hspost}@{}}%
\column{14}{@{}>{\hspre}l<{\hspost}@{}}%
\column{17}{@{}>{\hspre}l<{\hspost}@{}}%
\column{20}{@{}>{\hspre}l<{\hspost}@{}}%
\column{23}{@{}>{\hspre}l<{\hspost}@{}}%
\column{26}{@{}>{\hspre}l<{\hspost}@{}}%
\column{46}{@{}>{\hspre}c<{\hspost}@{}}%
\column{46E}{@{}l@{}}%
\column{49}{@{}>{\hspre}l<{\hspost}@{}}%
\column{58}{@{}>{\hspre}l<{\hspost}@{}}%
\column{67}{@{}>{\hspre}l<{\hspost}@{}}%
\column{72}{@{}>{\hspre}c<{\hspost}@{}}%
\column{72E}{@{}l@{}}%
\column{75}{@{}>{\hspre}l<{\hspost}@{}}%
\column{E}{@{}>{\hspre}l<{\hspost}@{}}%
\>[3]{}\Varid{cback}_4\mathrel{=}\Varid{foldr}\;\Varid{aux}\;\Varid{cpow}{}\<[E]%
\\
\>[3]{}\hsindent{2}{}\<[5]%
\>[5]{}\mathbf{where}\;{}\<[12]%
\>[12]{}\Varid{aux}\;{}\<[17]%
\>[17]{}\Varid{o}\;{}\<[20]%
\>[20]{}\Varid{k}\;{}\<[23]%
\>[23]{}\mathrm{0}\;{}\<[26]%
\>[26]{}\Varid{p}{}\<[46]%
\>[46]{}\mathrel{=}{}\<[46E]%
\>[49]{}\Varid{k}\;{}\<[58]%
\>[58]{}\Varid{p}\;{}\<[67]%
\>[67]{}\Varid{o}{}\<[E]%
\\
\>[12]{}\Varid{aux}\;{}\<[17]%
\>[17]{}\Varid{o}\;{}\<[20]%
\>[20]{}\Varid{k}\;{}\<[23]%
\>[23]{}\Varid{q}\;{}\<[26]%
\>[26]{}\mathrm{0}{}\<[46]%
\>[46]{}\mathrel{=}{}\<[46E]%
\>[49]{}\Varid{k}\;{}\<[58]%
\>[58]{}\mathrm{0}\;{}\<[67]%
\>[67]{}\Varid{o}{}\<[E]%
\\
\>[12]{}\Varid{aux}\;{}\<[17]%
\>[17]{}\Varid{o}\;{}\<[20]%
\>[20]{}\Varid{k}\;{}\<[23]%
\>[23]{}\Varid{q}\;{}\<[26]%
\>[26]{}\Varid{p}\mid \Varid{p}\mathbin{>}\mathrm{0}\mathrel{\wedge}\Varid{q}\mathbin{>}\mathrm{0}{}\<[46]%
\>[46]{}\mathrel{=}{}\<[46E]%
\>[49]{}\Varid{aux}\;\Varid{o}\;\Varid{k}\;{}\<[58]%
\>[58]{}(\Varid{q}\mathbin{-}\mathrm{1})\;{}\<[67]%
\>[67]{}(\Varid{p'}{}\<[72]%
\>[72]{}\mathbin{-}{}\<[72E]%
\>[75]{}\mathrm{1}){}\<[E]%
\\
\>[12]{}\hsindent{2}{}\<[14]%
\>[14]{}\mathbf{where}\;\Varid{p'}{}\<[46]%
\>[46]{}\mathrel{=}{}\<[46E]%
\>[49]{}\Varid{aux}\;\Varid{o}\;\Varid{k}\;{}\<[58]%
\>[58]{}\Varid{q}\;{}\<[67]%
\>[67]{}(\Varid{p}{}\<[72]%
\>[72]{}\mathbin{-}{}\<[72E]%
\>[75]{}\mathrm{1}){}\<[E]%
\ColumnHook
\end{hscode}\resethooks
The remaining steps are business as usual.  Eliminate \ensuremath{\Varid{p}} from \ensuremath{\Varid{aux}} by currying,
                   \begin{hscode}\SaveRestoreHook
\column{B}{@{}>{\hspre}l<{\hspost}@{}}%
\column{3}{@{}>{\hspre}l<{\hspost}@{}}%
\column{5}{@{}>{\hspre}l<{\hspost}@{}}%
\column{7}{@{}>{\hspre}l<{\hspost}@{}}%
\column{9}{@{}>{\hspre}l<{\hspost}@{}}%
\column{12}{@{}>{\hspre}l<{\hspost}@{}}%
\column{15}{@{}>{\hspre}l<{\hspost}@{}}%
\column{16}{@{}>{\hspre}l<{\hspost}@{}}%
\column{18}{@{}>{\hspre}l<{\hspost}@{}}%
\column{22}{@{}>{\hspre}l<{\hspost}@{}}%
\column{33}{@{}>{\hspre}c<{\hspost}@{}}%
\column{33E}{@{}l@{}}%
\column{36}{@{}>{\hspre}l<{\hspost}@{}}%
\column{45}{@{}>{\hspre}l<{\hspost}@{}}%
\column{54}{@{}>{\hspre}l<{\hspost}@{}}%
\column{59}{@{}>{\hspre}c<{\hspost}@{}}%
\column{59E}{@{}l@{}}%
\column{62}{@{}>{\hspre}l<{\hspost}@{}}%
\column{E}{@{}>{\hspre}l<{\hspost}@{}}%
\>[3]{}\Varid{cback}_5\mathrel{=}\Varid{foldr}\;\Varid{aux}\;\Varid{cpow}{}\<[E]%
\\
\>[3]{}\hsindent{2}{}\<[5]%
\>[5]{}\mathbf{where}{}\<[E]%
\\
\>[5]{}\hsindent{2}{}\<[7]%
\>[7]{}\Varid{aux}\;{}\<[12]%
\>[12]{}\Varid{o}\;{}\<[15]%
\>[15]{}\Varid{k}\;\mathrm{0}{}\<[33]%
\>[33]{}\mathrel{=}{}\<[33E]%
\>[36]{}\lambda \Varid{p}\to \Varid{k}\;\Varid{p}\;\Varid{o}{}\<[E]%
\\
\>[5]{}\hsindent{2}{}\<[7]%
\>[7]{}\Varid{aux}\;{}\<[12]%
\>[12]{}\Varid{o}\;{}\<[15]%
\>[15]{}\Varid{k}\;\Varid{q}\mid \Varid{q}\mathbin{>}\mathrm{0}{}\<[33]%
\>[33]{}\mathrel{=}{}\<[33E]%
\>[36]{}\Varid{aux2}{}\<[E]%
\\
\>[7]{}\hsindent{2}{}\<[9]%
\>[9]{}\mathbf{where}\;{}\<[16]%
\>[16]{}\Varid{aux2}\;{}\<[22]%
\>[22]{}\mathrm{0}{}\<[33]%
\>[33]{}\mathrel{=}{}\<[33E]%
\>[36]{}\Varid{k}\;{}\<[45]%
\>[45]{}\mathrm{0}\;{}\<[54]%
\>[54]{}\Varid{o}{}\<[E]%
\\
\>[16]{}\Varid{aux2}\;{}\<[22]%
\>[22]{}\Varid{p}\mid \Varid{p}\mathbin{>}\mathrm{0}{}\<[33]%
\>[33]{}\mathrel{=}{}\<[33E]%
\>[36]{}\Varid{aux}\;\Varid{o}\;\Varid{k}\;{}\<[45]%
\>[45]{}(\Varid{q}\mathbin{-}\mathrm{1})\;{}\<[54]%
\>[54]{}(\Varid{p'}{}\<[59]%
\>[59]{}\mathbin{-}{}\<[59E]%
\>[62]{}\mathrm{1}){}\<[E]%
\\
\>[16]{}\hsindent{2}{}\<[18]%
\>[18]{}\mathbf{where}\;\Varid{p'}{}\<[33]%
\>[33]{}\mathrel{=}{}\<[33E]%
\>[36]{}\Varid{aux}\;\Varid{o}\;\Varid{k}\;{}\<[45]%
\>[45]{}\Varid{q}\;{}\<[54]%
\>[54]{}(\Varid{p}{}\<[59]%
\>[59]{}\mathbin{-}{}\<[59E]%
\>[62]{}\mathrm{1}){}\<[E]%
\ColumnHook
\end{hscode}\resethooks
abstract from the ``productive'' self-reference by beta expansion,
                       \begin{hscode}\SaveRestoreHook
\column{B}{@{}>{\hspre}l<{\hspost}@{}}%
\column{3}{@{}>{\hspre}l<{\hspost}@{}}%
\column{5}{@{}>{\hspre}l<{\hspost}@{}}%
\column{7}{@{}>{\hspre}l<{\hspost}@{}}%
\column{9}{@{}>{\hspre}l<{\hspost}@{}}%
\column{12}{@{}>{\hspre}l<{\hspost}@{}}%
\column{15}{@{}>{\hspre}l<{\hspost}@{}}%
\column{16}{@{}>{\hspre}l<{\hspost}@{}}%
\column{18}{@{}>{\hspre}l<{\hspost}@{}}%
\column{22}{@{}>{\hspre}l<{\hspost}@{}}%
\column{25}{@{}>{\hspre}l<{\hspost}@{}}%
\column{36}{@{}>{\hspre}c<{\hspost}@{}}%
\column{36E}{@{}l@{}}%
\column{39}{@{}>{\hspre}l<{\hspost}@{}}%
\column{48}{@{}>{\hspre}l<{\hspost}@{}}%
\column{57}{@{}>{\hspre}l<{\hspost}@{}}%
\column{62}{@{}>{\hspre}c<{\hspost}@{}}%
\column{62E}{@{}l@{}}%
\column{65}{@{}>{\hspre}l<{\hspost}@{}}%
\column{E}{@{}>{\hspre}l<{\hspost}@{}}%
\>[3]{}\Varid{cback}_{5\mathrm{a}}\mathrel{=}\Varid{foldr}\;\Varid{aux}\;\Varid{cpow}{}\<[E]%
\\
\>[3]{}\hsindent{2}{}\<[5]%
\>[5]{}\mathbf{where}{}\<[E]%
\\
\>[5]{}\hsindent{2}{}\<[7]%
\>[7]{}\Varid{aux}\;{}\<[12]%
\>[12]{}\Varid{o}\;{}\<[15]%
\>[15]{}\Varid{k}\;\mathrm{0}{}\<[36]%
\>[36]{}\mathrel{=}{}\<[36E]%
\>[39]{}\lambda \Varid{p}\to \Varid{k}\;\Varid{p}\;\Varid{o}{}\<[E]%
\\
\>[5]{}\hsindent{2}{}\<[7]%
\>[7]{}\Varid{aux}\;{}\<[12]%
\>[12]{}\Varid{o}\;{}\<[15]%
\>[15]{}\Varid{k}\;\Varid{q}\mid \Varid{q}\mathbin{>}\mathrm{0}{}\<[36]%
\>[36]{}\mathrel{=}{}\<[36E]%
\>[39]{}\Varid{aux2}\;(\Varid{aux}\;\Varid{o}\;\Varid{k}\;(\Varid{q}\mathbin{-}\mathrm{1})){}\<[E]%
\\
\>[7]{}\hsindent{2}{}\<[9]%
\>[9]{}\mathbf{where}\;{}\<[16]%
\>[16]{}\Varid{aux2}\;{}\<[22]%
\>[22]{}\Varid{f}\;{}\<[25]%
\>[25]{}\mathrm{0}{}\<[36]%
\>[36]{}\mathrel{=}{}\<[36E]%
\>[39]{}\Varid{k}\;{}\<[48]%
\>[48]{}\mathrm{0}\;{}\<[57]%
\>[57]{}\Varid{o}{}\<[E]%
\\
\>[16]{}\Varid{aux2}\;{}\<[22]%
\>[22]{}\Varid{f}\;{}\<[25]%
\>[25]{}\Varid{p}\mid \Varid{p}\mathbin{>}\mathrm{0}{}\<[36]%
\>[36]{}\mathrel{=}{}\<[36E]%
\>[39]{}\Varid{f}\;{}\<[57]%
\>[57]{}(\Varid{p'}{}\<[62]%
\>[62]{}\mathbin{-}{}\<[62E]%
\>[65]{}\mathrm{1}){}\<[E]%
\\
\>[16]{}\hsindent{2}{}\<[18]%
\>[18]{}\mathbf{where}\;\Varid{p'}{}\<[36]%
\>[36]{}\mathrel{=}{}\<[36E]%
\>[39]{}\Varid{aux}\;\Varid{o}\;\Varid{k}\;{}\<[48]%
\>[48]{}\Varid{q}\;{}\<[57]%
\>[57]{}(\Varid{p}{}\<[62]%
\>[62]{}\mathbin{-}{}\<[62E]%
\>[65]{}\mathrm{1}){}\<[E]%
\ColumnHook
\end{hscode}\resethooks
and substitute
                       \begin{hscode}\SaveRestoreHook
\column{B}{@{}>{\hspre}l<{\hspost}@{}}%
\column{3}{@{}>{\hspre}l<{\hspost}@{}}%
\column{E}{@{}>{\hspre}l<{\hspost}@{}}%
\>[3]{}\Varid{aux}\;\Varid{o}\;\Varid{k}\;\Varid{q}\equiv \Varid{aux2}\;(\Varid{aux}\;\Varid{o}\;\Varid{k}\;(\Varid{q}\mathbin{-}\mathrm{1}))\equiv \Varid{aux2}\;\Varid{f}{}\<[E]%
\ColumnHook
\end{hscode}\resethooks
to obtain:
                       \begin{hscode}\SaveRestoreHook
\column{B}{@{}>{\hspre}l<{\hspost}@{}}%
\column{3}{@{}>{\hspre}l<{\hspost}@{}}%
\column{5}{@{}>{\hspre}l<{\hspost}@{}}%
\column{7}{@{}>{\hspre}l<{\hspost}@{}}%
\column{9}{@{}>{\hspre}l<{\hspost}@{}}%
\column{12}{@{}>{\hspre}l<{\hspost}@{}}%
\column{15}{@{}>{\hspre}l<{\hspost}@{}}%
\column{16}{@{}>{\hspre}l<{\hspost}@{}}%
\column{18}{@{}>{\hspre}l<{\hspost}@{}}%
\column{22}{@{}>{\hspre}l<{\hspost}@{}}%
\column{25}{@{}>{\hspre}l<{\hspost}@{}}%
\column{36}{@{}>{\hspre}c<{\hspost}@{}}%
\column{36E}{@{}l@{}}%
\column{39}{@{}>{\hspre}l<{\hspost}@{}}%
\column{49}{@{}>{\hspre}l<{\hspost}@{}}%
\column{58}{@{}>{\hspre}l<{\hspost}@{}}%
\column{63}{@{}>{\hspre}c<{\hspost}@{}}%
\column{63E}{@{}l@{}}%
\column{66}{@{}>{\hspre}l<{\hspost}@{}}%
\column{E}{@{}>{\hspre}l<{\hspost}@{}}%
\>[3]{}\Varid{cback}_{5\mathrm{b}}\mathrel{=}\Varid{foldr}\;\Varid{aux}\;\Varid{cpow}{}\<[E]%
\\
\>[3]{}\hsindent{2}{}\<[5]%
\>[5]{}\mathbf{where}{}\<[E]%
\\
\>[5]{}\hsindent{2}{}\<[7]%
\>[7]{}\Varid{aux}\;{}\<[12]%
\>[12]{}\Varid{o}\;{}\<[15]%
\>[15]{}\Varid{k}\;{}\<[18]%
\>[18]{}\mathrm{0}{}\<[36]%
\>[36]{}\mathrel{=}{}\<[36E]%
\>[39]{}\lambda \Varid{p}\to \Varid{k}\;\Varid{p}\;\Varid{o}{}\<[E]%
\\
\>[5]{}\hsindent{2}{}\<[7]%
\>[7]{}\Varid{aux}\;{}\<[12]%
\>[12]{}\Varid{o}\;{}\<[15]%
\>[15]{}\Varid{k}\;{}\<[18]%
\>[18]{}\Varid{q}\mid \Varid{q}\mathbin{>}\mathrm{0}{}\<[36]%
\>[36]{}\mathrel{=}{}\<[36E]%
\>[39]{}\Varid{aux2}\;(\Varid{aux}\;\Varid{o}\;\Varid{k}\;(\Varid{q}\mathbin{-}\mathrm{1})){}\<[E]%
\\
\>[7]{}\hsindent{2}{}\<[9]%
\>[9]{}\mathbf{where}\;{}\<[16]%
\>[16]{}\Varid{aux2}\;{}\<[22]%
\>[22]{}\Varid{f}\;{}\<[25]%
\>[25]{}\mathrm{0}{}\<[36]%
\>[36]{}\mathrel{=}{}\<[36E]%
\>[39]{}\Varid{k}\;{}\<[49]%
\>[49]{}\mathrm{0}\;{}\<[58]%
\>[58]{}\Varid{o}{}\<[E]%
\\
\>[16]{}\Varid{aux2}\;{}\<[22]%
\>[22]{}\Varid{f}\;{}\<[25]%
\>[25]{}\Varid{p}\mid \Varid{p}\mathbin{>}\mathrm{0}{}\<[36]%
\>[36]{}\mathrel{=}{}\<[36E]%
\>[39]{}\Varid{f}\;{}\<[58]%
\>[58]{}(\Varid{p'}{}\<[63]%
\>[63]{}\mathbin{-}{}\<[63E]%
\>[66]{}\mathrm{1}){}\<[E]%
\\
\>[16]{}\hsindent{2}{}\<[18]%
\>[18]{}\mathbf{where}\;\Varid{p'}{}\<[36]%
\>[36]{}\mathrel{=}{}\<[36E]%
\>[39]{}\Varid{aux2}\;\Varid{f}\;{}\<[58]%
\>[58]{}(\Varid{p}{}\<[63]%
\>[63]{}\mathbin{-}{}\<[63E]%
\>[66]{}\mathrm{1}){}\<[E]%
\ColumnHook
\end{hscode}\resethooks
Perform recursion abstraction in \ensuremath{\Varid{q}},
                       \begin{hscode}\SaveRestoreHook
\column{B}{@{}>{\hspre}l<{\hspost}@{}}%
\column{3}{@{}>{\hspre}l<{\hspost}@{}}%
\column{5}{@{}>{\hspre}l<{\hspost}@{}}%
\column{7}{@{}>{\hspre}l<{\hspost}@{}}%
\column{9}{@{}>{\hspre}l<{\hspost}@{}}%
\column{16}{@{}>{\hspre}l<{\hspost}@{}}%
\column{18}{@{}>{\hspre}l<{\hspost}@{}}%
\column{22}{@{}>{\hspre}l<{\hspost}@{}}%
\column{25}{@{}>{\hspre}l<{\hspost}@{}}%
\column{36}{@{}>{\hspre}c<{\hspost}@{}}%
\column{36E}{@{}l@{}}%
\column{39}{@{}>{\hspre}l<{\hspost}@{}}%
\column{49}{@{}>{\hspre}l<{\hspost}@{}}%
\column{58}{@{}>{\hspre}l<{\hspost}@{}}%
\column{63}{@{}>{\hspre}c<{\hspost}@{}}%
\column{63E}{@{}l@{}}%
\column{66}{@{}>{\hspre}l<{\hspost}@{}}%
\column{E}{@{}>{\hspre}l<{\hspost}@{}}%
\>[3]{}\Varid{cback}_6\mathrel{=}\Varid{foldr}\;\Varid{aux}\;\Varid{cpow}{}\<[E]%
\\
\>[3]{}\hsindent{2}{}\<[5]%
\>[5]{}\mathbf{where}{}\<[E]%
\\
\>[5]{}\hsindent{2}{}\<[7]%
\>[7]{}\Varid{aux}\;\Varid{o}\;\Varid{k}\mathrel{=}\Varid{foldn}\;\Varid{aux2}\;(\lambda \Varid{p}\to \Varid{k}\;\Varid{p}\;\Varid{o}){}\<[E]%
\\
\>[7]{}\hsindent{2}{}\<[9]%
\>[9]{}\mathbf{where}\;{}\<[16]%
\>[16]{}\Varid{aux2}\;{}\<[22]%
\>[22]{}\Varid{f}\;{}\<[25]%
\>[25]{}\mathrm{0}{}\<[36]%
\>[36]{}\mathrel{=}{}\<[36E]%
\>[39]{}\Varid{k}\;{}\<[49]%
\>[49]{}\mathrm{0}\;{}\<[58]%
\>[58]{}\Varid{o}{}\<[E]%
\\
\>[16]{}\Varid{aux2}\;{}\<[22]%
\>[22]{}\Varid{f}\;{}\<[25]%
\>[25]{}\Varid{p}\mid \Varid{p}\mathbin{>}\mathrm{0}{}\<[36]%
\>[36]{}\mathrel{=}{}\<[36E]%
\>[39]{}\Varid{f}\;{}\<[58]%
\>[58]{}(\Varid{p'}{}\<[63]%
\>[63]{}\mathbin{-}{}\<[63E]%
\>[66]{}\mathrm{1}){}\<[E]%
\\
\>[16]{}\hsindent{2}{}\<[18]%
\>[18]{}\mathbf{where}\;\Varid{p'}{}\<[36]%
\>[36]{}\mathrel{=}{}\<[36E]%
\>[39]{}\Varid{aux2}\;\Varid{f}\;{}\<[58]%
\>[58]{}(\Varid{p}{}\<[63]%
\>[63]{}\mathbin{-}{}\<[63E]%
\>[66]{}\mathrm{1}){}\<[E]%
\ColumnHook
\end{hscode}\resethooks
beta-expand the auxiliary expression \ensuremath{\Varid{p'}},
                       \begin{hscode}\SaveRestoreHook
\column{B}{@{}>{\hspre}l<{\hspost}@{}}%
\column{3}{@{}>{\hspre}l<{\hspost}@{}}%
\column{5}{@{}>{\hspre}l<{\hspost}@{}}%
\column{7}{@{}>{\hspre}l<{\hspost}@{}}%
\column{9}{@{}>{\hspre}l<{\hspost}@{}}%
\column{11}{@{}>{\hspre}l<{\hspost}@{}}%
\column{17}{@{}>{\hspre}l<{\hspost}@{}}%
\column{20}{@{}>{\hspre}l<{\hspost}@{}}%
\column{31}{@{}>{\hspre}c<{\hspost}@{}}%
\column{31E}{@{}l@{}}%
\column{34}{@{}>{\hspre}l<{\hspost}@{}}%
\column{E}{@{}>{\hspre}l<{\hspost}@{}}%
\>[3]{}\Varid{cback}_7\mathrel{=}\Varid{foldr}\;\Varid{aux}\;\Varid{cpow}{}\<[E]%
\\
\>[3]{}\hsindent{2}{}\<[5]%
\>[5]{}\mathbf{where}{}\<[E]%
\\
\>[5]{}\hsindent{2}{}\<[7]%
\>[7]{}\Varid{aux}\;\Varid{o}\;\Varid{k}\mathrel{=}\Varid{foldn}\;\Varid{aux2}\;(\lambda \Varid{p}\to \Varid{k}\;\Varid{p}\;\Varid{o}){}\<[E]%
\\
\>[7]{}\hsindent{2}{}\<[9]%
\>[9]{}\mathbf{where}{}\<[E]%
\\
\>[9]{}\hsindent{2}{}\<[11]%
\>[11]{}\Varid{aux2}\;{}\<[17]%
\>[17]{}\Varid{f}\;{}\<[20]%
\>[20]{}\mathrm{0}{}\<[31]%
\>[31]{}\mathrel{=}{}\<[31E]%
\>[34]{}\Varid{k}\;\mathrm{0}\;\Varid{o}{}\<[E]%
\\
\>[9]{}\hsindent{2}{}\<[11]%
\>[11]{}\Varid{aux2}\;{}\<[17]%
\>[17]{}\Varid{f}\;{}\<[20]%
\>[20]{}\Varid{p}\mid \Varid{p}\mathbin{>}\mathrm{0}{}\<[31]%
\>[31]{}\mathrel{=}{}\<[31E]%
\>[34]{}(\lambda \Varid{p'}\to \Varid{f}\;(\Varid{p'}\mathbin{-}\mathrm{1}))\;(\Varid{aux2}\;\Varid{f}\;(\Varid{p}\mathbin{-}\mathrm{1})){}\<[E]%
\ColumnHook
\end{hscode}\resethooks
and perform recursion abstraction in \ensuremath{\Varid{p}}:
\begin{hscode}\SaveRestoreHook
\column{B}{@{}>{\hspre}l<{\hspost}@{}}%
\column{3}{@{}>{\hspre}l<{\hspost}@{}}%
\column{5}{@{}>{\hspre}l<{\hspost}@{}}%
\column{7}{@{}>{\hspre}l<{\hspost}@{}}%
\column{9}{@{}>{\hspre}l<{\hspost}@{}}%
\column{16}{@{}>{\hspre}l<{\hspost}@{}}%
\column{E}{@{}>{\hspre}l<{\hspost}@{}}%
\>[3]{}\Varid{cback}_8\mathrel{=}\Varid{foldr}\;\Varid{aux}\;\Varid{cpow}{}\<[E]%
\\
\>[3]{}\hsindent{2}{}\<[5]%
\>[5]{}\mathbf{where}{}\<[E]%
\\
\>[5]{}\hsindent{2}{}\<[7]%
\>[7]{}\Varid{aux}\;\Varid{o}\;\Varid{k}\mathrel{=}\Varid{foldn}\;\Varid{aux2}\;(\lambda \Varid{p}\to \Varid{k}\;\Varid{p}\;\Varid{o}){}\<[E]%
\\
\>[7]{}\hsindent{2}{}\<[9]%
\>[9]{}\mathbf{where}\;{}\<[16]%
\>[16]{}\Varid{aux2}\;\Varid{f}\mathrel{=}\Varid{foldn}\;(\lambda \Varid{p'}\to \Varid{f}\;(\Varid{p'}\mathbin{-}\mathrm{1}))\;(\Varid{k}\;\mathrm{0}\;\Varid{o}){}\<[E]%
\ColumnHook
\end{hscode}\resethooks

\noindent Final cosmetic translation for improved point-freeness:
\begin{hscode}\SaveRestoreHook
\column{B}{@{}>{\hspre}l<{\hspost}@{}}%
\column{3}{@{}>{\hspre}l<{\hspost}@{}}%
\column{5}{@{}>{\hspre}l<{\hspost}@{}}%
\column{7}{@{}>{\hspre}l<{\hspost}@{}}%
\column{9}{@{}>{\hspre}l<{\hspost}@{}}%
\column{13}{@{}>{\hspre}l<{\hspost}@{}}%
\column{16}{@{}>{\hspre}l<{\hspost}@{}}%
\column{17}{@{}>{\hspre}l<{\hspost}@{}}%
\column{20}{@{}>{\hspre}c<{\hspost}@{}}%
\column{20E}{@{}l@{}}%
\column{23}{@{}>{\hspre}l<{\hspost}@{}}%
\column{E}{@{}>{\hspre}l<{\hspost}@{}}%
\>[3]{}\Varid{cback}_9\mathrel{=}\Varid{foldr}\;\Varid{aux}\;\Varid{cpow}{}\<[E]%
\\
\>[3]{}\hsindent{2}{}\<[5]%
\>[5]{}\mathbf{where}{}\<[E]%
\\
\>[5]{}\hsindent{2}{}\<[7]%
\>[7]{}\Varid{aux}\;{}\<[13]%
\>[13]{}\Varid{o}\;{}\<[16]%
\>[16]{}\Varid{k}{}\<[20]%
\>[20]{}\mathrel{=}{}\<[20E]%
\>[23]{}\Varid{foldn}\;\Varid{aux2}\;(\Varid{flip}\;\Varid{k}\;\Varid{o}){}\<[E]%
\\
\>[7]{}\hsindent{2}{}\<[9]%
\>[9]{}\mathbf{where}\;{}\<[17]%
\>[17]{}\Varid{aux2}\;\Varid{f}\mathrel{=}\Varid{foldn}\;(\Varid{f}\mathbin{\circ}(-\;\mathrm{1}))\;(\Varid{k}\;\mathrm{0}\;\Varid{o}){}\<[E]%
\ColumnHook
\end{hscode}\resethooks
\subsection{Verification}               

Verification is a more complex issue here because of the sheer number
of cases in indirections.  Since no significant new insights are to be
gained, we give just one case, corresponding to the fourth equation of
\ensuremath{\Varid{conway}_0}, for illustration purposes:
\begin{hscode}\SaveRestoreHook
\column{B}{@{}>{\hspre}l<{\hspost}@{}}%
\column{3}{@{}>{\hspre}l<{\hspost}@{}}%
\column{31}{@{}>{\hspre}c<{\hspost}@{}}%
\column{31E}{@{}l@{}}%
\column{35}{@{}>{\hspre}l<{\hspost}@{}}%
\column{37}{@{}>{\hspre}l<{\hspost}@{}}%
\column{44}{@{}>{\hspre}l<{\hspost}@{}}%
\column{47}{@{}>{\hspre}l<{\hspost}@{}}%
\column{80}{@{}>{\hspre}l<{\hspost}@{}}%
\column{E}{@{}>{\hspre}l<{\hspost}@{}}%
\>[3]{}\Varid{cfront}_2\;\Varid{cback}_8\;(\mathrm{1}\mathbin{:}\Varid{p}\mathbin{:}\Varid{x}){}\<[31]%
\>[31]{}\equiv {}\<[31E]%
\>[35]{}\Varid{cfront}_1\;\Varid{cback}_8\;(\Varid{map}\;(-\;\mathrm{1})\;(\mathrm{1}\mathbin{:}\Varid{p}\mathbin{:}\Varid{x})){}\<[E]%
\\
\>[31]{}\equiv {}\<[31E]%
\>[35]{}\Varid{cfront}_1\;\Varid{cback}_8\;(\mathrm{0}\mathbin{:}\Varid{p}\mathbin{-}\mathrm{1}\mathbin{:}\Varid{x'}){}\<[E]%
\\
\>[35]{}\hsindent{2}{}\<[37]%
\>[37]{}\mathbf{where}\;\Varid{x'}\mathrel{=}\Varid{map}\;(-\;\mathrm{1})\;\Varid{x}{}\<[E]%
\\
\>[31]{}\equiv {}\<[31E]%
\>[35]{}\Varid{cback}_8\;\Varid{x'}\;\mathrm{0}\;(\Varid{p}\mathbin{-}\mathrm{1}){}\<[E]%
\\
\>[31]{}\equiv {}\<[31E]%
\>[44]{}\Varid{foldr}\;\Varid{aux}\;\Varid{cpow}\;\Varid{x'}\;\mathrm{0}\;(\Varid{p}\mathbin{-}\mathrm{1}){}\<[E]%
\\
\>[31]{}\equiv {}\<[31E]%
\>[35]{}\Varid{aux}\;\Varid{o}\;({}\<[44]%
\>[44]{}\Varid{foldr}\;\Varid{aux}\;\Varid{cpow}\;\Varid{y})\;\mathrm{0}\;(\Varid{p}\mathbin{-}\mathrm{1}){}\<[E]%
\\
\>[35]{}\hsindent{2}{}\<[37]%
\>[37]{}\mathbf{where}\;(\Varid{o}\mathbin{:}\Varid{y})\mathrel{=}\Varid{x'}{}\<[E]%
\\
\>[31]{}\equiv {}\<[31E]%
\>[35]{}\Varid{foldn}\;\Varid{aux2}\;{}\<[47]%
\>[47]{}(\lambda \Varid{p}\to \Varid{foldr}\;\Varid{aux}\;\Varid{cpow}\;\Varid{y}\;\Varid{p}\;\Varid{o})\;\mathrm{0}\;{}\<[80]%
\>[80]{}(\Varid{p}\mathbin{-}\mathrm{1}){}\<[E]%
\\
\>[31]{}\equiv {}\<[31E]%
\>[47]{}(\lambda \Varid{p}\to \Varid{foldr}\;\Varid{aux}\;\Varid{cpow}\;\Varid{y}\;\Varid{p}\;\Varid{o})\;{}\<[80]%
\>[80]{}(\Varid{p}\mathbin{-}\mathrm{1}){}\<[E]%
\\
\>[31]{}\equiv {}\<[31E]%
\>[35]{}\Varid{foldr}\;\Varid{aux}\;\Varid{cpow}\;\Varid{y}\;(\Varid{p}\mathbin{-}\mathrm{1})\;\Varid{o}{}\<[E]%
\\
\>[31]{}\equiv {}\<[31E]%
\>[35]{}\Varid{cback}_8\;\Varid{y}\;(\Varid{p}\mathbin{-}\mathrm{1})\;\Varid{o}{}\<[E]%
\\
\>[31]{}\equiv {}\<[31E]%
\>[35]{}\Varid{cfront}_1\;\Varid{cback}_8\;(\Varid{p}\mathbin{-}\mathrm{1}\mathbin{:}\Varid{o}\mathbin{:}\Varid{y}){}\<[E]%
\\
\>[31]{}\equiv {}\<[31E]%
\>[35]{}\Varid{cfront}_1\;\Varid{cback}_8\;(\Varid{p}\mathbin{-}\mathrm{1}\mathbin{:}\Varid{x'}){}\<[E]%
\\
\>[31]{}\equiv {}\<[31E]%
\>[35]{}\Varid{cfront}_1\;\Varid{cback}_8\;(\Varid{map}\;(-\;\mathrm{1})\;(\Varid{p}\mathbin{:}\Varid{x})){}\<[E]%
\\
\>[31]{}\equiv {}\<[31E]%
\>[35]{}\Varid{cfront}_2\;\Varid{cback}_8\;(\Varid{p}\mathbin{:}\Varid{x}){}\<[E]%
\ColumnHook
\end{hscode}\resethooks
$\Box$
               
\subsection{Analysis}           
               
The type analysis of the recursion scheme of Conway's chained arrows
is analogous to the preceding examples.  The third, outermost layer of folding has yet a more complex underlying datatype.
\begin{hscode}\SaveRestoreHook
\column{B}{@{}>{\hspre}l<{\hspost}@{}}%
\column{3}{@{}>{\hspre}l<{\hspost}@{}}%
\column{5}{@{}>{\hspre}l<{\hspost}@{}}%
\column{7}{@{}>{\hspre}l<{\hspost}@{}}%
\column{9}{@{}>{\hspre}l<{\hspost}@{}}%
\column{16}{@{}>{\hspre}l<{\hspost}@{}}%
\column{E}{@{}>{\hspre}l<{\hspost}@{}}%
\>[3]{}\Varid{foldr}_3\mathbin{::}\Conid{FoldR}\;\Conid{Integer}\;(\Conid{Integer}\to \Conid{Integer}\to \Conid{Integer}){}\<[E]%
\\
\>[3]{}\Varid{foldr}_3\mathrel{=}\Varid{foldr}{}\<[E]%
\\[\blanklineskip]%
\>[3]{}\Varid{cback}_{10}\mathrel{=}\Varid{foldr}_3\;\Varid{aux}\;\Varid{cpow}{}\<[E]%
\\
\>[3]{}\hsindent{2}{}\<[5]%
\>[5]{}\mathbf{where}{}\<[E]%
\\
\>[5]{}\hsindent{2}{}\<[7]%
\>[7]{}\Varid{aux}\;\Varid{o}\;\Varid{k}\mathrel{=}\Varid{foldn}_2\;\Varid{aux2}\;(\Varid{flip}\;\Varid{k}\;\Varid{o}){}\<[E]%
\\
\>[7]{}\hsindent{2}{}\<[9]%
\>[9]{}\mathbf{where}\;{}\<[16]%
\>[16]{}\Varid{aux2}\;\Varid{f}\mathrel{=}\Varid{foldn}_1\;(\Varid{f}\mathbin{\circ}(-\;\mathrm{1}))\;(\Varid{k}\;\mathrm{0}\;\Varid{o}){}\<[E]%
\ColumnHook
\end{hscode}\resethooks

\pagebreak
\section{Conclusion}
\label{conclusion}

Recursion abstraction is a sophisticated program transformation
technique.  Its straightforward applications are at the heart of the
algebra-of-programs approach to functional programming.  It is a
matter of taste whether definitions in terms of recursion operators
are more intuitive and readable than definitions in terms of
self-reference.  But without doubt, the primitive forms lends
themselves more easily to formal, in particular equational, reasoning.

Iterated recursion abstraction is notably more tricky than just a
single step.  Self-referen\-ces need to be eliminated by clever
context-sensitive beta expansion.  The Ackermann function is, as for
many other questions, just the right introductory example to teach the
technique.  In particular, its argument order leads the way naturally.
Conway's chained arrow notation, on the other hand, is a significantly
more difficult nut to crack, and is rather at the high end of
demonstrations for the potential of the technique.

The central trick has now been used four times in three exercises, so the general pattern should have become apparent.  In summary, a multiply nested self-referential cycle in a definition is broken by bringing the function into the form
\begin{hscode}\SaveRestoreHook
\column{B}{@{}>{\hspre}l<{\hspost}@{}}%
\column{3}{@{}>{\hspre}l<{\hspost}@{}}%
\column{5}{@{}>{\hspre}l<{\hspost}@{}}%
\column{6}{@{}>{\hspre}l<{\hspost}@{}}%
\column{12}{@{}>{\hspre}l<{\hspost}@{}}%
\column{17}{@{}>{\hspre}c<{\hspost}@{}}%
\column{17E}{@{}l@{}}%
\column{20}{@{}>{\hspre}l<{\hspost}@{}}%
\column{E}{@{}>{\hspre}l<{\hspost}@{}}%
\>[3]{}\Varid{h}\;{}\<[6]%
\>[6]{}\mathrm{0}{}\<[17]%
\>[17]{}\mathrel{=}{}\<[17E]%
\>[20]{}\Varid{e}{}\<[E]%
\\
\>[3]{}\Varid{h}\;{}\<[6]%
\>[6]{}\Varid{n}\mid \Varid{n}\mathbin{>}\mathrm{0}{}\<[17]%
\>[17]{}\mathrel{=}{}\<[17E]%
\>[20]{}\Varid{g}\;(\Varid{h}\;(\Varid{n}\mathbin{-}\mathrm{1})){}\<[E]%
\\
\>[3]{}\hsindent{2}{}\<[5]%
\>[5]{}\mathbf{where}\;{}\<[12]%
\>[12]{}\Varid{g}\;\Varid{f}\mathrel{=}\Varid{i}\;(\Varid{h}\;\Varid{n}){}\<[E]%
\ColumnHook
\end{hscode}\resethooks
where the auxiliary higher-order function \ensuremath{\Varid{i}} typically arises
``virtually'' by beta abstraction from a given expression in which \ensuremath{\Varid{h}\;\Varid{n}} occurs.  Then one proceeds to
\begin{hscode}\SaveRestoreHook
\column{B}{@{}>{\hspre}l<{\hspost}@{}}%
\column{3}{@{}>{\hspre}l<{\hspost}@{}}%
\column{5}{@{}>{\hspre}l<{\hspost}@{}}%
\column{6}{@{}>{\hspre}l<{\hspost}@{}}%
\column{17}{@{}>{\hspre}c<{\hspost}@{}}%
\column{17E}{@{}l@{}}%
\column{20}{@{}>{\hspre}l<{\hspost}@{}}%
\column{E}{@{}>{\hspre}l<{\hspost}@{}}%
\>[3]{}\Varid{h}\;{}\<[6]%
\>[6]{}\mathrm{0}{}\<[17]%
\>[17]{}\mathrel{=}{}\<[17E]%
\>[20]{}\Varid{e}{}\<[E]%
\\
\>[3]{}\Varid{h}\;{}\<[6]%
\>[6]{}\Varid{n}\mid \Varid{n}\mathbin{>}\mathrm{0}{}\<[17]%
\>[17]{}\mathrel{=}{}\<[17E]%
\>[20]{}\Varid{g'}\;(\Varid{h}\;(\Varid{n}\mathbin{-}\mathrm{1})){}\<[E]%
\\
\>[3]{}\hsindent{2}{}\<[5]%
\>[5]{}\mathbf{where}\;\Varid{g'}\;\Varid{f}\mathrel{=}\Varid{i}\;(\Varid{g'}\;\Varid{f}){}\<[E]%
\ColumnHook
\end{hscode}\resethooks
Globally, \ensuremath{\Varid{g}} and \ensuremath{\Varid{g'}} are quite different functions, but we find for \ensuremath{\Varid{n}\mathbin{>}\mathrm{0}}:
\begin{hscode}\SaveRestoreHook
\column{B}{@{}>{\hspre}l<{\hspost}@{}}%
\column{3}{@{}>{\hspre}l<{\hspost}@{}}%
\column{7}{@{}>{\hspre}l<{\hspost}@{}}%
\column{31}{@{}>{\hspre}c<{\hspost}@{}}%
\column{31E}{@{}l@{}}%
\column{35}{@{}>{\hspre}l<{\hspost}@{}}%
\column{42}{@{}>{\hspre}l<{\hspost}@{}}%
\column{E}{@{}>{\hspre}l<{\hspost}@{}}%
\>[3]{}\Varid{g}\;{}\<[7]%
\>[7]{}(\Varid{h}\;(\Varid{n}\mathbin{-}\mathrm{1}))\equiv \Varid{i}\;(\Varid{h}\;\Varid{n}){}\<[31]%
\>[31]{}\equiv {}\<[31E]%
\>[35]{}\Varid{i}\;(\Varid{g}\;{}\<[42]%
\>[42]{}(\Varid{h}\;(\Varid{n}\mathbin{-}\mathrm{1}))){}\<[E]%
\\
\>[3]{}\Varid{g'}\;{}\<[7]%
\>[7]{}(\Varid{h}\;(\Varid{n}\mathbin{-}\mathrm{1})){}\<[31]%
\>[31]{}\equiv {}\<[31E]%
\>[35]{}\Varid{i}\;(\Varid{g'}\;{}\<[42]%
\>[42]{}(\Varid{h}\;(\Varid{n}\mathbin{-}\mathrm{1}))){}\<[E]%
\ColumnHook
\end{hscode}\resethooks
That is, both \ensuremath{\Varid{g}\;\Varid{f}} and \ensuremath{\Varid{g'}\;\Varid{f}}, where \ensuremath{\Varid{f}\mathrel{=}\Varid{h}\;(\Varid{n}\mathbin{-}\mathrm{1})}, are fixed
points of \ensuremath{\Varid{i}}.  Since we are in a functional programming language with
unique fixed point semantics as the very foundation of
self-referential definitions, we may substitute one for the other.
Then the outer recursion can be reduced to primitive form
\begin{hscode}\SaveRestoreHook
\column{B}{@{}>{\hspre}l<{\hspost}@{}}%
\column{3}{@{}>{\hspre}l<{\hspost}@{}}%
\column{5}{@{}>{\hspre}l<{\hspost}@{}}%
\column{E}{@{}>{\hspre}l<{\hspost}@{}}%
\>[3]{}\Varid{h}\mathrel{=}\Varid{foldn}\;\Varid{e}\;\Varid{g'}{}\<[E]%
\\
\>[3]{}\hsindent{2}{}\<[5]%
\>[5]{}\mathbf{where}\;\Varid{g'}\;\Varid{f}\mathrel{=}\Varid{i}\;(\Varid{g'}\;\Varid{f}){}\<[E]%
\ColumnHook
\end{hscode}\resethooks
and the inner self-reference of \ensuremath{\Varid{g'}} can be processed further by recursion abstraction, using either the same or more basic techniques.
               
It seems plausible that this technique will work for a large class of multiply nested recursive functions.  The auxiliary tactics that we have employed, namely moving nested recursion to final argument position and refactoring deep pattern matches, are possibly also more general heuristics, and merit further investigation.

\subsection{Outlook}
               
We close by posing several open problems as challenges to the reader:

\begin{enumerate}
\item Assess whether the typical, relative ease of inductive reasoning about recursive functions in primitive form carries over to well-known, albeit non-trivial identities concerning our example functions, such as:
\begin{hscode}\SaveRestoreHook
\column{B}{@{}>{\hspre}l<{\hspost}@{}}%
\column{3}{@{}>{\hspre}l<{\hspost}@{}}%
\column{E}{@{}>{\hspre}l<{\hspost}@{}}%
\>[3]{}\Varid{ack}\;(\Varid{m}\mathbin{+}\mathrm{2})\;\Varid{n}\equiv \Varid{knuth}\;\mathrm{2}\;\Varid{m}\;(\Varid{n}\mathbin{+}\mathrm{3})\mathbin{-}\mathrm{3}{}\<[E]%
\ColumnHook
\end{hscode}\resethooks
\item As an important special case of equational reasoning, demonstrate program simplifications, such as \emph{fusion} rules, for expressions involving our example functions.
\item Use the higher-order primitive recursion framework as a toolkit for synthesizing novel functions with interesting, recursive equational definitions.
\end{enumerate}

\section*{Acknowledgments}
               
Anonymous referees of a previous manuscript have made valuable contributions to the evolution of this article.
               
\bibliographystyle{abbrv}
\bibliography{primitives}

\end{document}